\documentclass[referee]{mn2e}
\usepackage{amsmath}

\usepackage{epsfig,rotate,graphicx}
\newcommand{\be}{\begin{equation}}
\newcommand{\ee}{\end{equation}}

\title[Multiple systems of super--Earths]{On the dynamics of multiple systems of hot
super--Earths and Neptunes:  Tidal circularization, resonance and the
HD~40307 system}

\author[Papaloizou and Terquem]{John C. B. Papaloizou$^1$\thanks{E-mail: J.C.B.Papaloizou@damtp.cam.ac.uk} and Caroline Terquem$^{2,3}$\thanks{E-mail: caroline.terquem@iap.fr}  \\
$^1$ Department of Applied Mathematics and Theoretical Physics,
University of Cambridge, Centre for Mathematical Sciences,\\
\  \ Wilberforce  Road, Cambridge, CB3 0WA, UK \\
$^2$ Institut d'Astrophysique de Paris, UPMC Univ Paris 06, CNRS,
UMR7095, 98 bis bd Arago, F-75014, Paris, France \\
$^3$ Institut Universitaire de France }

\begin{document}

\maketitle

\begin{abstract}

In this paper, we consider the dynamics of
  a system of hot super--Earths or Neptunes such as HD~40307.  
 We  show that, as  tidal interaction
with the central star leads to small eccentricities, the planets in this system could be 
undergoing  resonant coupling even though 
 the period ratios depart significantly 
from very precise commensurability.   In a three planet system, this is indicated by
 the fact that  resonant angles librate
or are associated with long term changes to the orbital elements. 
In HD~40307, we expect that three resonant angles could be involved in this way.
 We propose that the planets in this system
 were in a strict Laplace resonance while they migrated through the disc.  
After entering the disc inner cavity, tidal interaction would cause the period ratios to increase from two
but with the inner pair deviating less than the outer pair, counter to what occurs in HD~40307.

However,  the relationship between these pairs that occurs in HD~40307  might be produced if the resonance is impulsively modified
by an event like a close encounter  shortly after the planetary system decouples from the disc.
 We find this to be in principle possible for
a small  relative perturbation  on the order  of  a  few $\times 10^{-3}$ but  then  we find  that  evolution to the present system in a reasonable time is possible
only if the masses are significantly larger 
than the minimum masses and tidal dissipation is  very effective.
 On the other hand we found that  a system like HD~40307 with minimum
masses and more realistic tidal dissipation could be produced if 
the eccentricity of the outermost planet was impulsively increased to $\sim 0.15.$  
 
We remark that
the form of resonantly coupled tidal evolution we consider here is quite general  and could be of
greater significance for  systems with inner planets 
on significantly shorter orbital periods characteristic of for example  CoRoT~7~b.

\end{abstract}


\begin{keywords}
planetary systems: formation --- planetary systems:
protoplanetary discs
\end{keywords}

\section{Introduction}
\label{sec:intro}

Neptune mass extrasolar planets around main sequence stars were first detected five years ago.  Since then, 27 planets with a projected mass lower than 25 earth masses have been discovered, the lightest one having a projected mass of 1.94~M$_{\oplus}$.  Among these objects, 17 have a semi--major axis smaller than 0.1~astronomical unit (au).  They are called hot Neptunes or hot super--Earths.   So far, two multiple planet systems with at least two such objects have been observed.   One of them is a four planet system around the M--dwarf GJ~581 (Bonfils et al. 2005, Udry et al. 2007, Mayor et al. 2009a).  The projected masses of the planets are 1.9, 15.6, 5.4 and 7.1~M$_{\oplus}$ and the periods are  3.15, 5.37, 12.93 and 66.8 days, respectively.  
Note that although the eccentricities of the two innermost planets are compatible with zero, those of the next  outermost and outermost  planets
 are 0.16 and 0.18 respectively.
 The second multiple system which we shall focus upon in this paper is that around the K--dwarf HD~40307 
(Mayor et al. 2009b).   It comprises  three 
planets with projected masses of 4.2, 6.9 and 9.1~M$_{\oplus}$ and periods of  4.31, 9.62 and 20.46 days, respectively.    
The eccentricities are all compatible with zero.   

If we label the planets in a system with successive integers
starting from the innermost labelled, 1, and moving outwards, then, in the system Gl~581, 
the ratio of the periods of planets 2 and 1 is 2.41 and that of the periods of planets 3 and 2 is 1.70. 
 We note that these numbers depart from 5/2 and 5/3 only by 4\% and 2\%, respectively.a
 In the HD~40307 system , the ratio of the periods of planets 2 and 1 is 2.23 and that of 
 planets 3 and 2 is 2.13, which depart from 2 by 11.5\% and 6.5\%, respectively. 
 Because departure from exact mean motion resonances in these systems is significant,
 {the importance of resonances for their dynamical evolution}  has been ruled out (Mayor et al. 2009a, 2009b, Barnes et al. 2009).
We shall address this aspect for the HD~40307 system  in this paper.

Migration due to tidal interaction with the disc is probably the mechanism by which 
planets end up on short period  orbits, as {\em in situ} formation requires very massive discs (e.g., Raymond et al. 2008).  Planet--planet scatterings may also lead to short period orbits, but only when 
tidal circularization is efficient.   According to Kennedy \& Kenyon (2008), scatterings are not a likely way of producing hot super--Earths, because of the long circularization timescales involved. 
In a previous paper (Terquem \& Papaloizou 2007, see also 
Brunini \& Cionco 2005),  we proposed a scenario for 
forming hot super--Earths on near commensurable orbits, in which 
a population of cores that 
assembled at some distance from the central star migrated inwards due
to tidal interaction with the disc while merging on their way in. 
We found that  evolution of an ensemble of cores in a disc almost always
led to a system of a few planets, with masses that depended on the
total initial mass of the system, on short period orbits with mean motions that frequently
exhibited near commensurabilities and, for long enough tidal
circularization times, apsidal lines that were locked together.
Starting with a population of 10 to 25 planets of 0.1 or 1~M$_\oplus$,
we ended up with typically between two and five planets with masses of a
few tenths of an earth mass or a few earth masses
 inside the disc inner edge.  Interaction with the
central star led to the initiation of tidal circularization of the orbits which,
together with possible close scatterings and final mergers, tended to
disrupt mean--motion resonances that were established during the
migration phase.  The system, however, often remained in a
configuration in which the orbital periods were close to
commensurability.  Apsidal locking of the orbits, if established
during migration, was often maintained through the action of these
processes.

This scenario has been questioned (Mayor et al. 2009a, 2009b) because the multiple planet systems of hot super--Earths detected so far do not exhibit close enough mean motion commensurabilities.   In this paper, we argue that {\em the system around HD~40307 does actually exhibit  the effects of resonances} through  secular effects produced by the action of the resonant angles
coupled with the action of tides from the central star, and that such a configuration could be produced if the system formed as described in Terquem \& Papaloizou (2007), but with the addition of relatively  small perturbations to  the orbits arising from close encounters
or collisions  after the system enters the disc inner cavity. 
The reason that what are regarded as resonant effects can arise even when departures from strict commensurability
are apparently large is that tidal circularization produces small eccentricities which, for first order resonances,  can be consistent with resonant
angle libration under those conditions (see Murray \& Dermott 1999).

We consider in detail the evolution of this system and other putative similar systems with scaled masses and orbital periods
after they form  a strict three planet Laplace resonance as a result of convergent inward orbital migration.
We go on to consider the onset of tidal circularization and how it leads to a separation of the semi-major axes and consequent increasing
departure from commensurability. We find that if the strict Laplace resonance is broken by a fairly small relative  perturbation
 corresponding to a few parts in a thousand,
continuing evolution in a modified form of the three planet resonance could in principle in the case of HD~40307 lead to a system
like the observed one.  However, this would require masses significantly larger than the minimum estimates and possibly
unrealistically efficient tidal dissipation with the tidal dissipation parameter $Q' \sim 10.$ 
However,  weaker  tidal effects could be responsible for some of the deviation
from strict commensurability, with the rest being produced by a larger perturbation resulting from the processes mentioned  above
inducing eccentricities of the order $0.1.$  
 Note however that  tidal  evolution
could  play a more significant role  in  similar  systems
with  shorter orbital periods.  

In section~\ref{sec:model}, we describe the numerical model we use to simulate the evolution of a system of three
 planets migrating in a disc and evolving under the tidal interaction with the star after they enter the disc inner cavity.  In section~\ref{sec:HD40307}, we simulate the system around HD~40307 using either the minimum or twice the minimum masses for the planets
under varying assumptions about initial eccentricities and 
 the strength of the tidal interaction.  We show that three of the four  possible resonant angles
associated with 2:1 commensurabilities either   librate or have long term time averages,
 indicating evolution of the system towards a resonant state driven by the tidal influence of the star.
 
 In a related appendix, we  give a semi--analytic model for a system in a three planet resonance undergoing 
circularization  with departures  from  strict mean motion commensurabilities. 
We note in passing that the discussion may also be applied
     to a two planet system near a 2:1 resonance.  We show that  the departure from commensurability
     increases with time being  $\propto t^{1/3}$ and derive an expression for the timescale required to attain a given departure that can be used to interpret {extrapolate from}  and generalize the numerical simulations. 
    
      In section~\ref{sec:simulations}, we present results of numerical simulations of a three planet system migrating in a disc, establishing a 4:2:1 Laplace like resonance, and evolving under tidal effects
     induced by  the star after entering the disc inner cavity. 
       As expected, as the orbits are being tidally circularized, the planets move away from exact commensurabilites, and the period ratios of the outer and inner pairs of planets increases. 
       
        The system is still resonant in the sense that some of the resonant angles continue to librate. 
  To get a larger period ratio for the inner pair of planets, as in HD~40307, some disruption of this Laplace resonant state is needed.
           In our simulations, this is achieved by  applying an   impulse (that could  result from  an encounter or collision) to the system.
This leads to departures  from exact commensurabilites of the form seen in HD~40307 while still retaining
            libration of  three of the four  resonant angles.  
         Finally, in section~\ref{sec:discussion} we discuss  and summarize our results.

\section{Model and initial conditions}
\label{sec:model}

The model we use has been described in Terquem \& Papaloizou (2007).  We
recall here the main features.  We consider a system consisting of a
primary star and N~planets embedded in a gaseous disc surrounding it
(for HD~40307, N =3).
The planets undergo gravitational interaction with each other and the
star and are acted on by tidal forces from the disc and the star.  The
evolution of the system is studied by means of the solution of  $N$--body problems,
in which the tidal interactions are included as dissipative forces.

The equations of motion are:
\begin{equation}
{d^2 {\bf r}_i\over dt^2} = -{GM_\star{\bf r}_i\over |{\bf r}_i|^3}
-\sum_{j=1\ne i}^N {Gm_j  \left({\bf r}_i-{\bf r}_j \right) \over |{\bf
    r}_i-{\bf r}_j |^3} -{\bf \Gamma} +{\bf \Gamma}_{i} +{\bf \Gamma}_{r} \; ,
\label{emot}
\end{equation}

\noindent where $M_\star$, $m_j$ and ${\bf r}_j$ denote the mass of
the central star, that of planet~$j$ and the position vector of planet
$j$, respectively.  The acceleration of the coordinate system based on
the central star (indirect term) is:
\begin{equation}
{\bf \Gamma}= \sum_{j=1}^N {Gm_j{\bf r}_{j} \over |{\bf r}_{j}|^3},
\label{indt}
\end{equation}

\noindent and that due to tidal interaction with the disc and/or the
star is dealt with through the addition of extra forces as in
Papaloizou~\& Larwood~(2000):
\begin{equation}
{\bf \Gamma}_{i} = -\frac{1}{t_{m,i}} \frac{d {\bf r}_i}{dt} -
\frac{2}{|{\bf r}_i|^2 t_{e,i}} \left( \frac{d {\bf r}_i}{dt} \cdot
{\bf r}_i \right) {\bf r}_i - \frac{2}{ t_{i,i}}
\left( \frac{d {\bf r}_i}{dt} \cdot {\bf e}_z \right) {\bf e}_z,
\label{Gammai}
\end{equation}

\noindent where $t_{m,i}$, $t_{e,i}$ and $t_{i,i}$ are the timescales
over which, respectively, the angular momentum, the eccentricity and
the inclination with respect to the unit normal ${\bf e}_z$ to the  assumed fixed gas
disc midplane change.  Evolution of the angular momentum and
inclination is due to tidal interaction with the disc, whereas
evolution of the eccentricity occurs due to both tidal interaction
with the disc and the star.  We have:
\begin{equation}
\frac{1}{t_{e,i}} = \frac{1}{t_{e,i}^d} + \frac{1}{t_{e,i}^s} ,
\end{equation}

\noindent where $t_{e,i}^d$ and $t_{e,i}^s$ are the contribution from
the disc and tides raised by the star, respectively.  Relativistic
effects are included through ${\bf \Gamma}_{r}$ ( see Papaloizou \&
Terquem 2001).

The way it is implemented, interaction with the star does not modify the
angular momentum of a single orbit (the orbital decay timescale due to
tidal interaction with the star, which  as indicated below 
is estimated to be  much longer than the
circularization timescale, has been ignored here).  However, in the
formulation above, eccentricity damping causes radial velocity damping,
which results in energy loss at constant angular momentum.  As a
consequence, both the semi--major axis and eccentricity are reduced
together (at least for a single planet).

\subsection {Orbital circularization due to tides from the central star}
\label{sec:startide}

The circularization timescale due to tidal interaction with the star
can be estimated, from expressions given by  Goldreich~\& Soter (1966), to be:
\begin{equation}
t_{e,i}^s = 4.065 \times 10^{4} \; \left( \frac{{\rm M}_\oplus}{m_i}
\right)^{2/3}  \left( \frac{20 a_i}{{\rm 1~au}} \right)^{6.5}  Q'
\; \; \; {\rm years} ,
\label{teccs}
\end{equation}

\noindent where $a_i$ is the semi--major axis of planet~$i.$ Here we
have adopted a mass density of 1~g~cm$^3$ for the planets 
(uncertainties in this quantity could be incorporated into a redefinition of $Q'$).
 The parameter
$Q'= 3Q/(2k_2),$ where $Q$ is the tidal dissipation function and $k_2$
is the Love number.   Equation (\ref{teccs}) applies to the situation
where tides raised by the central star are dissipated in the planet in the limit
of small eccentricity which is appropriate here.
Tides raised by the planet in the star have  been  estimated to be unimportant
for the low mass planets of interest (eg. Barnes et al 2009). 
 For solar system planets in the terrestrial mass
range, Goldreich \& Soter (1966) give estimates for $Q$ in the range
10--500 and $k_2 \sim 0.3$, which correspond to $Q'$ in the range
50--2500.  Of course this parameter must be 
regarded as being very uncertain for extrasolar planets, and accordingly
we have considered values of  $Q'$ in the range $10-1000$ in this paper. 
 For an earth mass planet at 0.05~au and $Q'=50$, we get
$t_{e,i}^s=2 \times 10^6$~years. This indicates that circularization processes
will be important for close in extrasolar planets in this mass range
over the entire range of $Q'$ considered.

But note that although circularization operates on each orbit
according to equation (\ref{teccs}), because the planets interact
they may enter into a decay mode with a decay timescale that is not  straightforwardly
estimated as above (see section \ref{initecc}) . Nonetheless  the above estimate is found to be indicative.

We have pointed out above that, in our formulation, tidal interaction
with the star does not change the angular momentum of a single orbit but
modifies the semi--major axis. The physical basis for this is that the planets
rapidly attain pseudo-synchronization (e.g., Ivanov \& Papaloizou 2007),   after which
they cannot store significant angular momentum  changes  through modifying
their  intrinsic  angular momenta.  

We restrict consideration for the moment to a single planet.
Since the orbital angular momentum is
conserved, $(da_i/dt)/a_i \simeq - 2e_i^2/t_{e,i}^s$ for small
eccentricities, where $e_i$ is the eccentricity of the planet we
consider.  Therefore $a_i$ changes only by about 5~percent in  $10^{10}$~years
for $Q'=10, $ $m_i=1$~M$_{\oplus}$  and $e_i=10^{-3}.$
However,  larger changes may occur for larger planet masses.  We investigate
the dependence of this form of evolution on the system configuration in detail below.

Finally we remark that all the simulation results reported here
were obtained assuming that $Q'$ is the same for all the planets.
However, tests were carried out with this assumption relaxed 
and these are alluded to in section \ref{sec:simulations} below.

\subsection{Type~I migration}\label{TypeImigration}

In the local linear treatment of type~I migration (e.g., Tanaka et
al.~2002), if the planet is not in contact with the disc, there is no
interaction between them so that $t_{m,i}$, $t_{e,i}^d$ and $t_{i,i}$
are taken to be infinite.  When the planet is in contact with the
disc, disc--planet interactions occur leading to orbital migration as
well as eccentricity and inclination damping (e.g., Ward~1997).  In
that case, away from the disc edge, for some simulations we have  adopted:
\begin{equation}
t_{m,i} = 146.0  \; \left[ 1+ \left( \frac{ e_i }{1.3 H/r}\right)^5 \right]
\left[ 1- \left( \frac{ e_i }{1.1 H/r}\right)^4 \right]^{-1}
\; \left( \frac{H/r}{0.05}
\right)^2 \;
\frac{{\rm M}_\odot}{M_d} \;
\frac{{\rm M}_\oplus}{m_i} \; \frac{a_i}{{\rm 1~au}} \; \; \; {\rm years} ,
\label{tm}
\end{equation}
\begin{equation}
t_{e,i}^d = 0.362 \; \left[ 1+ 0.25 \left( \frac{ e_i }{H/r}\right)^3
\right] \; \left( \frac{H/r}{0.05}
\right)^4 \;
\frac{{\rm M}_\odot}{M_d} \; \frac{{\rm M}_\oplus}{m_i} \;
\frac{a_i}{{\rm 1~au}} \; \; \; {\rm years} ,
\label{te}
\end{equation}

\noindent and $t_{i,i}=t_{e,i}$ (eq.~[31] and~[32] of Papaloizou~\&
Larwood~2000 with $f_s=0.6$).  Here $e_i$ is the eccentricity of
planet~$i$, $H/r$ is the disc aspect ratio and $M_d$ is the disc mass
contained within 5~au.  We have assumed here that the disc surface
mass density varies like $r^{-3/2}$.  For a 1~earth mass planet on a
quasi--circular orbit at 1~au, we get $t_{m,i} \sim 10^5$~yr and
$t_{e,i}^d \sim 500$~yr for $M_d=10^{-3}$~M$_\odot$ and $H/r=0.05$.
The timescales given by equations~(\ref{tm}) and~(\ref{te})
can be used not only for small values of $e_i$, but also for
eccentricities larger than $H/r$.

We have also carried out simulations with 
$t_{m,i}$ and $t_{e,i}^d$ taken to be constants and thus independent of eccentricity.
The commensurabilities that are formed in the system depend
on the ratio of the   migration rate to the orbital frequency,
with  commensurabilities of low order and low degree forming when this ratio is small.
Therefore, using equations~(\ref{tm}) and~(\ref{te}) or constant timescales 
give very similar results provided the eccentricity damping
is ultimately effective  and the magnitudes of the migration rates
are comparable in both cases at zero eccentricity.

 The type~I migration rates
to be used are uncertain  even when the surface density
is known,  largely because of uncertainties regarding the 
effectiveness of coorbital torques (e.g., Paardekooper \& Melema 2006, 
Pardekooper \& Papaloizou 2008, 2009). 
 In this context there are indications from modelling the observational data
that the adopted type I migration rate should be significantly
below that predicted by (\ref{tm}) and~(\ref{te}) or Tanaka et al (2002)
(see Schlaufman et al 2009). 
Hence we explore a range of possible values.


\section{HD~40307 and three planet resonance}
\label{sec:HD40307}

Three super--Earths have been detected  orbiting the
star HD~40307  through the radial velocity variations 
they produce (Mayor et al. 2009a).  The planets have projected masses of
4.2, 6.9 and 9.2~M$_{\earth}$ and the reported semi--major axes are 
0.047, 0.081 and
0.134~au, respectively, which correspond to
period ratios of 2.26
and 2.13 for the inner and the outer pairs of planets, respectively.    
The periods reported by Mayor et al. (2009a) are
4.31, 9.62 and 20.46~days, which correspond to period ratios of 2.23  and 2.13.
 The derived
eccentricities are compatible with zero. 
We will refer to the planets as~1, 2 and 3, planet~1 being the innermost object.

Because
the period ratios differ significantly from 2, resonances have been ruled out
(Mayor et al. 2009a, Barnes et al. 2009).   However, the period ratios  are not far from 4:2:1, being  characteristic of a
 three planet resonance.  This indicates  that  such a resonance might play
a role at some evolutionary phase of the system.  To  investigate this hypothesis, we
have simulated the system by means of $N$--body calculations  
using the method described in the previous section. 
The equations of motion are integrated using the Bulirsch--Stoer
method (e.g., Press et al.~1993).

The planets were set up on circular
orbits around a 0.77~M$_{\odot}$ star  (the estimated mass of HD~40307) with  both the 
minimum masses
indicated above and also masses a factor of two larger.
The initial semi--major axes were 
taken to be those indicated above,
except in two  simulations described below, where 
the innermost semi--major axis was adjusted 
to make the period  ratio of the innermost pair $2.23$ instead of 2.26.

\begin{figure}
\includegraphics[width=16cm]{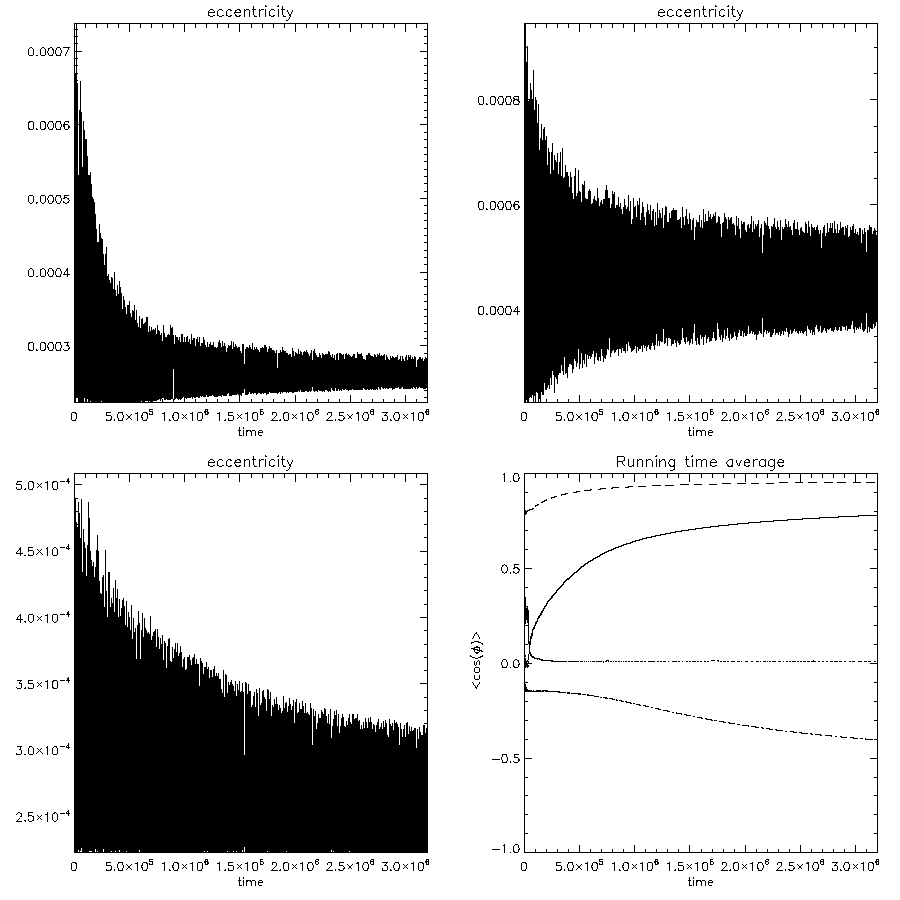} 
\caption{  Time dependent evolution for  Q'=10 and minimum masses.
Shown is the eccentricity of the innermost planet ({\em upper left panel}), middle planet ({\em upper right panel}) and outermost planet  ({\em lower right panel}) as a function of time.
Running time averages
of the cosines of the  resonant angles are shown in
the {\em lower right panel}. These correspond to
$\Phi_1=2\lambda_2-\lambda_1-\varpi_1$(full line),
$\Phi_2=2\lambda_2-\lambda_1-\varpi_2$( dotted line),
$\Phi_3=2\lambda_3-\lambda_2-\varpi_2$(dashed  line),
$\Phi_4=2\lambda_3-\lambda_2-\varpi_3$( dot dashed line).
For all plots time is measured in yr.}
 \label{figjp1}
\end{figure}

 The  results of  a calculation with the tidal parameter
$Q'=10$ and minimum masses  are displayed in
Fig.~\ref{figjp1}, which shows the evolution of the eccentricities
and runnimg time averages  
of the cosines of the resonant angles 
$\Phi_1= 2\lambda_2-\lambda_1-\varpi_1,$
$\Phi_2 = 2\lambda_2-\lambda_1-\varpi_2,$
$\Phi_3 = 2\lambda_3-\lambda_2-\varpi_2$ and 
$\Phi_4 = 2\lambda_3-\lambda_2-\varpi_3$ 
 over a timescale of 3.6~million years.
Here $\lambda_i$  and $\tilde{\omega}_i$ are the mean longitude 
and the argument of
pericentre of planet $i$, respectively.
 The semi--major axes do not vary significantly  with time
while the eccentricities   remain small,  below $10^{-3}$.  However,
tidal circularization processes are important for this run as the eccentricity 
of the outermost planet and the amplitude of the oscillations of the eccentricity of the 
inner two planets tends to decrease.  Interestingly, the  running time averages of the 
cosines of the resonant
angles are not zero, as would be expected for a non resonant system.
They are such that the values for $\cos \Phi_1$ and $\cos \Phi_3$ approach unity,
 while the value for $\cos \Phi_4$ moves towards $-1$, corresponding to the
 angle approaching  $\pi.$
The angle $\Phi_2$ exhibits non resonant behaviour.

\begin{figure}
\includegraphics[width=16cm]{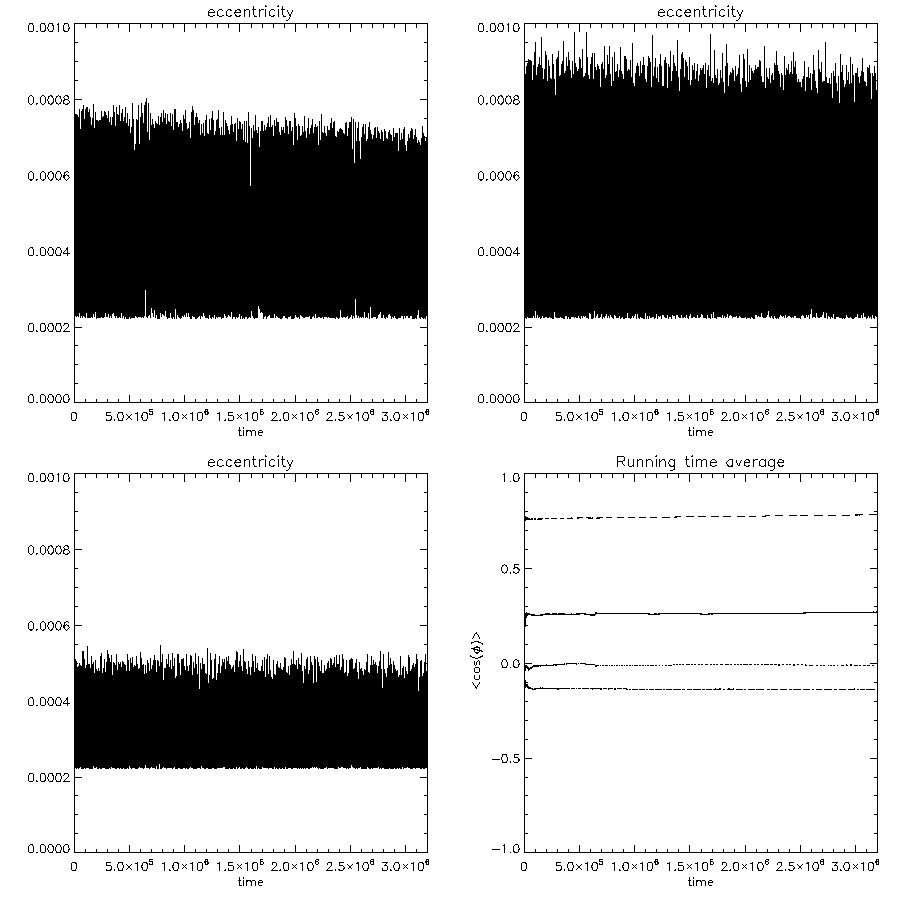} \caption{  As in  Fig.\ref{figjp1}  but for $Q'=1000$,
 minimum masses and initial innermost   period ratio 2.23.}
 \label{figjp2}
\end{figure}

 We have also investigated the dependence of the evolution
on $Q',$ small variations of the initial semi-major axes
and the magnitudes of the planet masses.  
Figure~\ref{figjp2} shows the evolution for $Q'=1000$,
minimum masses and the initial innermost semimajor axis
adjusted to make the innermost period ratio  2.23. In this case,
the mean values of the eccentricities show evidence only of limited evolution.
Nonetheless, running time averages of the cosines of $\Phi_1$  $\Phi_3$ and
 $\Phi_4$ quickly attain steady non zero values. This is consistent with the evolution
of these runs being driven by tidal effects and towards a resonant 
state in which the resonant angles are associated with secular evolution
of the  orbital elements. This is able to  occur without precise commensurability
only because the eccentricities are very small (see appendix A).
For comparison in Fig. \ref{figjp2a} we plot the running means of 
$e_1\cos\Phi_1,$ $e_2\cos\Phi_3,$ and  $e_3\cos\Phi_4$
for the same run. These also attain non zero values  also indicating 
long term variations  associated  with the corresponding  resonant angles.  Note  that the cosines
of the resonant angles are not special in this respect (see also appendix A). 

The dependence  on the planet mass scale is illustrated
in Fig.~\ref{figjp3} which shows the evolution for  $Q'=10$,
twice the minimum masses  and without adjustment of the innermost
semi-major axis. Figure \ref{figjp4} gives the evolution for the same
system but with the initial innermost semimajor axis
adjusted to make the innermost period ratio 2.23.  Both these runs show similar evolution
to  the minimum mass case illustrated in 
Fig.~\ref{figjp1} but scaled to  larger eccentricities. 
Thus  the results are not sensitive to the choice of minimum masses or small 
changes in the semi-major axes.

\begin{figure}
\begin{center}
\includegraphics[width=16cm]{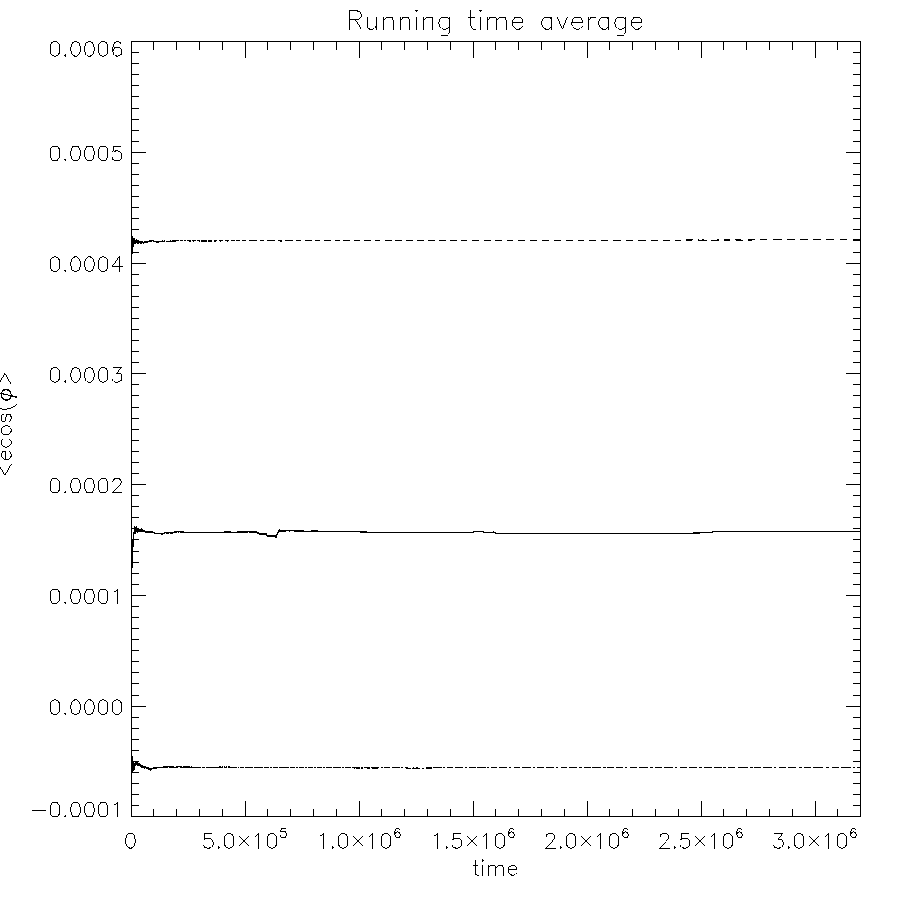}
\caption{  Evolution of the running time
averages of $e_1\cos\Phi_1$ (solid line)
$e_2\cos\Phi_3$ (dashed line) and $e_3\cos\Phi_4$ (dot-dashed  line) for  Q'=1000 and minimum masses.
Time is measured in yr.}
 \label{figjp2a}
\end{center}
\end{figure}

\begin{figure}
\includegraphics[width=16cm]{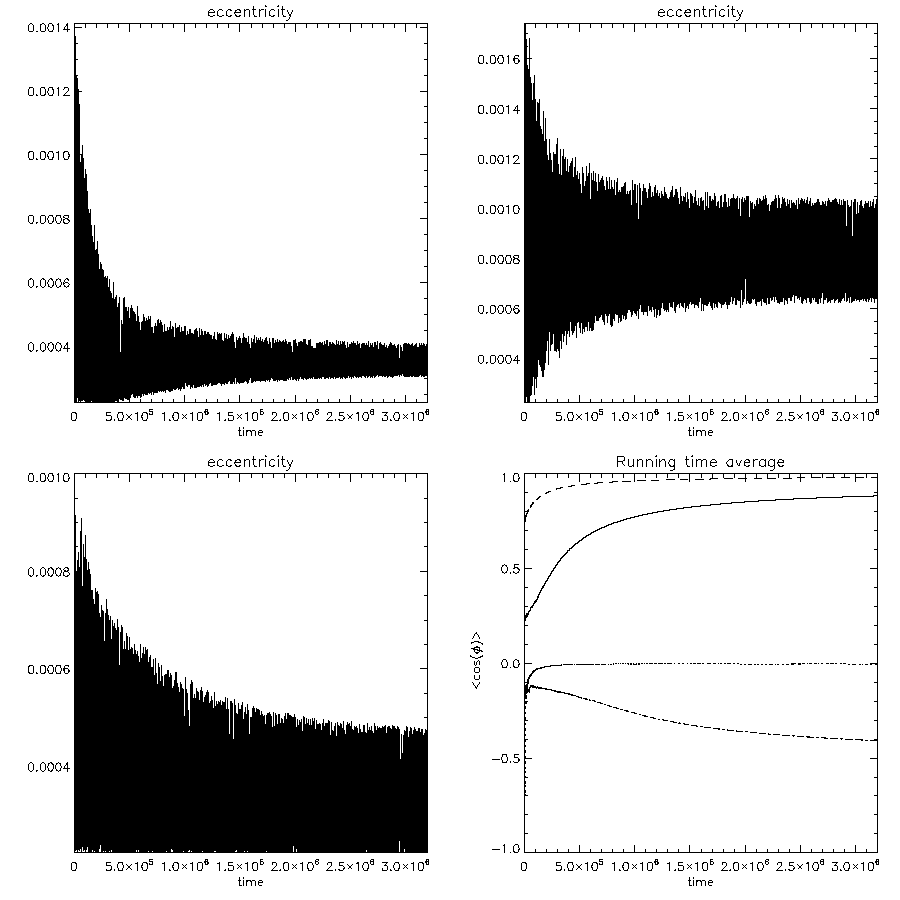} \caption{ As in  Fig.\ref{figjp1}
but with  Q'=10
and with  masses a factor of two larger than the minimum.}
 \label{figjp3}
\end{figure}

\begin{figure}
\includegraphics[width=16cm]{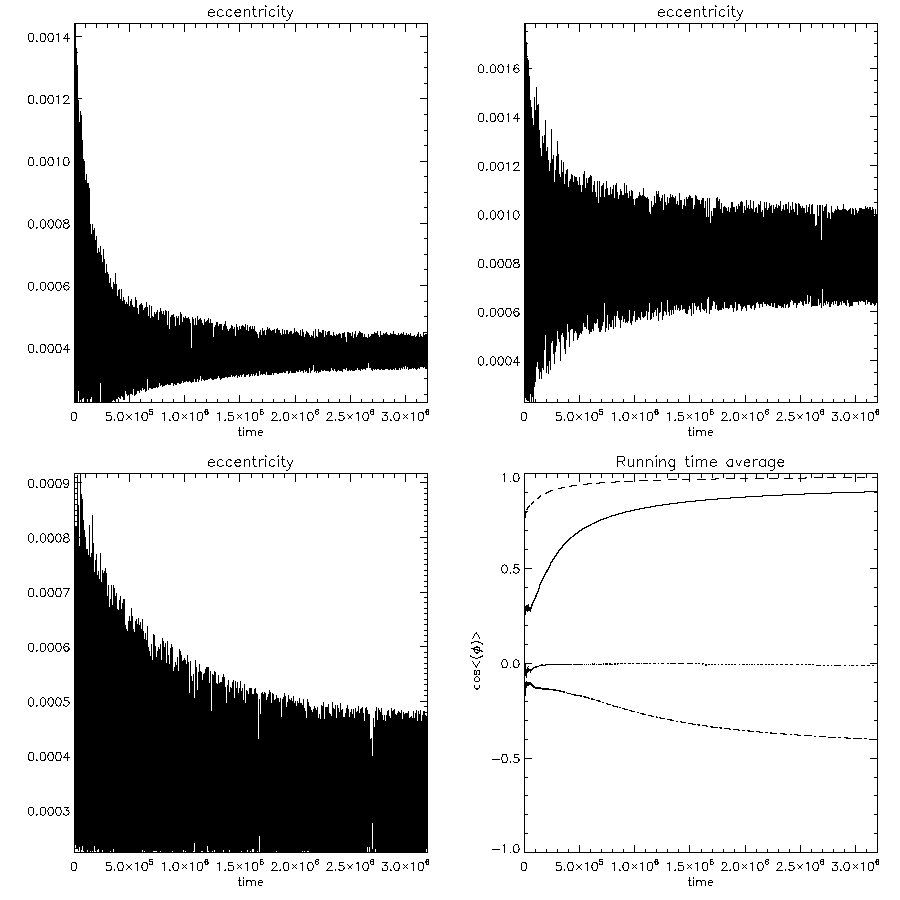} \caption{ As in  Fig.\ref{figjp3} 
  but with initial inner period ratio 2.23.}
 \label{figjp4}
\end{figure}

\begin{figure}
\begin{center}
\includegraphics[width=16cm]{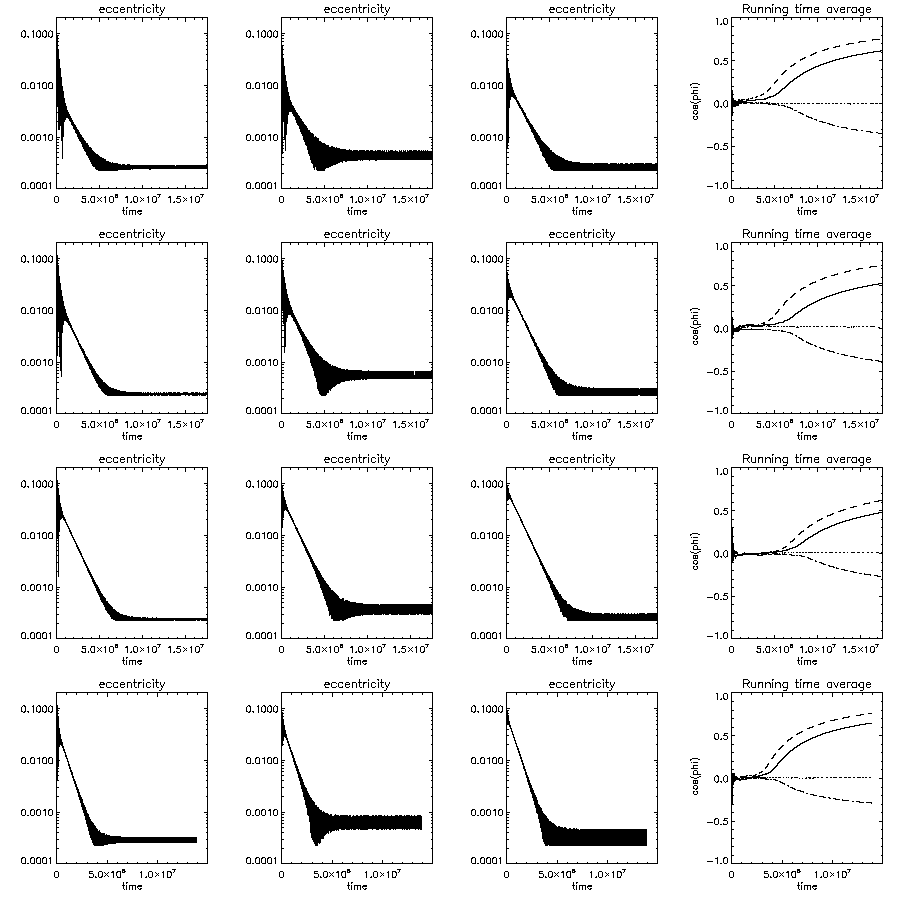}
\caption{   The evolution of the eccentricities  and running time averages of the resonant angles
are plotted as functions of time measured in years for simulations with $Q'=10.$
The first  second and third rows of panels correspond to the cases where the inner, middle
and outer planets were started with an initial eccentricity (see text) respectively
and minimum masses were adopted.
The fourth row is the same as the third except that the planet masses were twice the minimum values. 
From left  to right for each row, the first panel shows the eccentricity
of the innermost planet, the second the eccentricity of the middle
planet and the third panel shows the eccentricity of the outermost planet.
The fourth panel shows running time averages of  $\cos(\Phi_1)$ (solid line),
 $\cos(\Phi_2)$ (dotted line)  
 $\cos(\Phi_3)$ (dashed line) and $\cos(\Phi_4)$ (dot-dashed  line).}
 \label{figjp6}
\end{center}
\end{figure}

\begin{figure}
\begin{center}
\includegraphics[width=16cm]{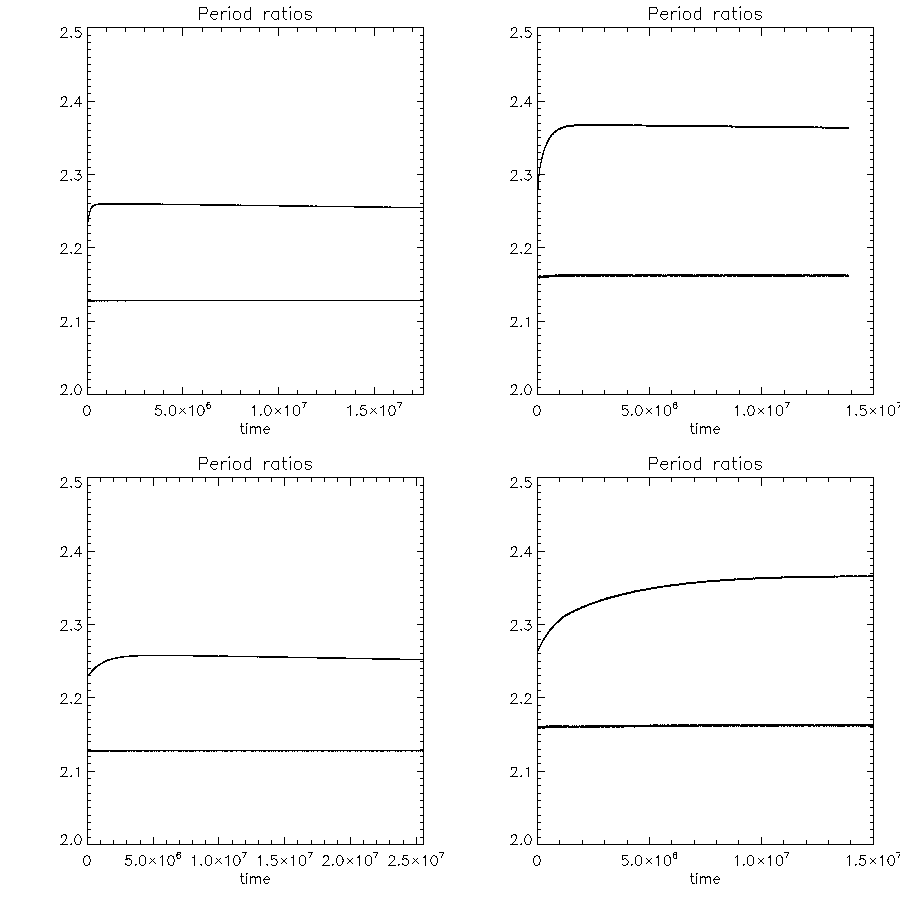}
\caption{ 
Evolution of the 
period ratios. The upper curve in each panel corresponds to the period ratio of the inner pair of planets
and the lower curve to the period ratio of the outer planets. 
The upper left panel corresponds to
minimum masses, $Q'=10$ and the innermost planet was
started with a non zero eccentricity (see text).
In the upper right panel the masses were twice the minimum
values,  $Q'=10$ and the outermost planet was started with a non zero 
eccentricity.  The lower left panel was as for the upper left panel
but $Q'=100.$  The lower right panel was  as for the upper right panel
but $Q'=100.$
Time is measured in yr.
Note that the shifts occurring in the period ratios
are independent of $Q'.$}
 \label{figjp7}
\end{center}
\end{figure}

\begin{figure}
\begin{center}
\includegraphics[width=16cm]{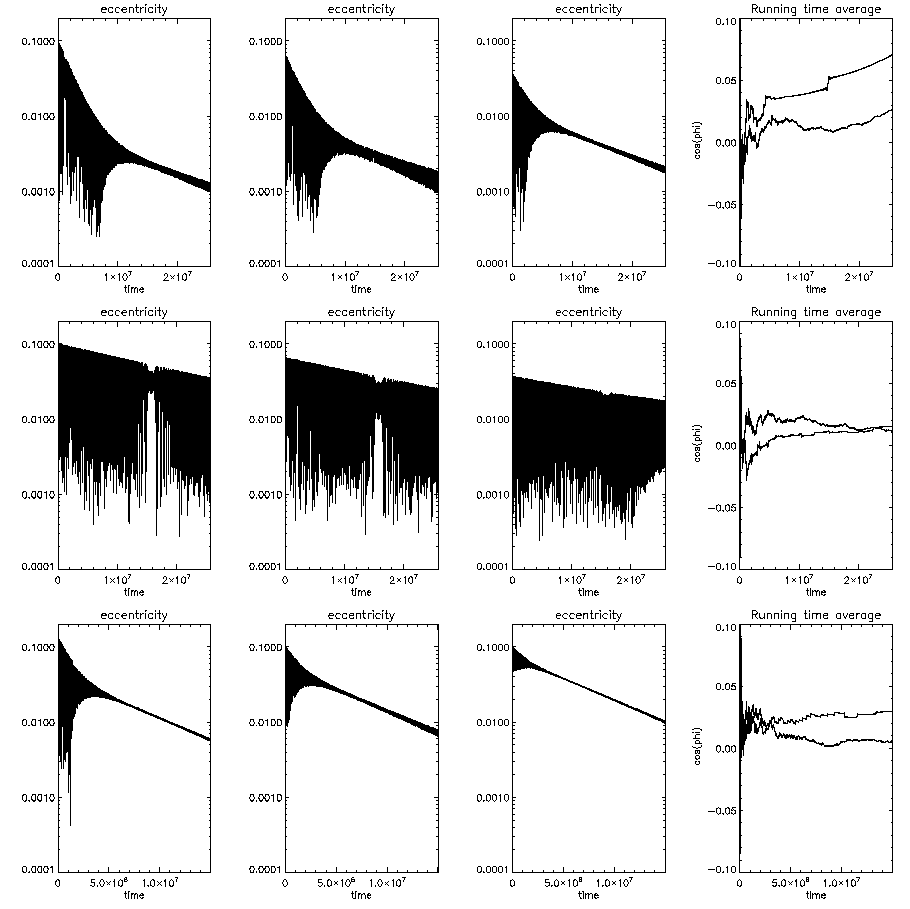}
\caption{ The evolution of the eccentricities  and running time averages of 
the cosines of the  resonant angles
are plotted as functions of time measured in years.
For the first  and second rows of panels,  $Q'=100$ and $Q'=1000$ respectively,
minimum masses were adopted  and  the innermost 
 planet was started with a non zero eccentricity (see text).
For the fourth  $Q'=100$  and the masses were twice the minimum values.
In this case the outermost planet was started with a non zero eccentricity.
From left  to right for each row, the first panel shows the eccentricity
of the innermost planet, the second the eccentricity of the middle
planet and the third panel shows the eccentricity of the outermost planet.
The fourth panel shows running time averages of   $\cos(\Phi_1)$   and
 $\cos(\Phi_3)$ (ultimately uppermost).}
 \label{figjp8}
\end{center}
\end{figure}

\subsection{The effect of initial eccentricity}\label{initecc}
 The calculations described above started with the planets
on circular orbits. The motivation for this is that it is expected 
that tidal circularization will  lead to a state closely approaching this
provided that initial eccentricities are not so large as to disrupt the system.
In order to explore this aspect in more detail, we have carried out a number of
simulations which started with non circular orbits.  We have explored
initial eccentricities $\le\sim 0.1$ and found that the system enters into an eccentricity  decay
mode that  causes the system to evolve towards  the
state we found starting from circular orbits in which three of the resonant angles
move towards libration with non zero values for the running time averages
of their cosines.

In order to illustrate this we describe here results obtained when just one of the
planets was started with a non zero eccentricity for various values
of $Q' \le 1000.$ To initiate the system it was started as for the simulation
illustrated in Figure  \ref{figjp1}  but with one of the planets
set  at pericentre  in  an orbit with the same angular momentum  but with
an eccentricity $e_{imp} \sim 0.1.$  
This is equivalent to transfer  to a less tightly bound  orbit with the original specific
angular momentum,  but with the specific energy multiplied by $1-e_{imp}^2.$
In all cases the eccentricities of the planets become coupled
and start to decay. After a while the system attains a slow  decay mode
for  which, in a time average sense,  the eccentricities are in an approximately constant 
ratio such that $e_2/e_1 \sim 1.2$ and $e_3/e_1 \sim 1.6.$

 Fig. \ref{figjp6} illustrates the evolution of the eccentricities  and
 running time averages of the cosines of the resonant angles
for minimum mass simulations with $Q'=10.$ Cases for which the innermost, middle
and outermost planet   respectively had an  initial eccentricity  are shown.
For comparison a case with twice the minimum masses
and for which the outermost planet was started
with a non zero eccentricity is also shown.
The evolution of all of these systems was as described above.

 However, the introduction of the initial eccentricity while conserving
the angular momentum of the system  
causes the semi-major axes and hence the period ratios in the system to change.
This evolution involves energy dissipation at fixed angular momentum
and thus the system must spread radially as  for an accretion disc
(eg. Lynden-Bell \& Pringle 1974). This causes the period ratios to increase
(see Fig. \ref{figjp7}).  When this happens the system attains period ratios that
differ from the original ones. However, this is easily compensated for
by starting with appropriately shifted initial conditions. 
The important point is that we get the same type of evolution independently
of such shifts. Fig. \ref{figjp7} shows that the largest shift occurs when the outermost
planet starts with an initial eccentricity. The period ratio of the innermost pair
 shifts from $2.23$  to $2.37.$ A consequence of this is that the forced eccentricity
 of the innermost planet is $\sim  25\%$ smaller in this case than in the case where the
 inner planet started with a non zero eccentricity (see Fig. \ref{figjp6}).
 We remark that the same evolution is obtained when the  masses are increased
 by a factor of two  except that the ultimate eccentricities are  a  factor between $1.5 $ and $2$ larger.
 The above trends are in qualitative agreement with a  simplified discussion of the resonant
 interaction between the planets that we give in  appendix A  even though
 only two of the resonant angles $\Phi_1$ and $\Phi_3$ are found to be librating
 at this point  with $\cos(\Phi_4)$ having a non zero  time average
 of relatively small magnitude.

Finally we have checked that changing $Q'$ gives the same form of solution
but with an evolutionary time that scales with $Q'.$
Results for  $Q'=100$ and $Q'=1000$  that confirm this view 
when compared with results already described for $Q'=10$ are illustrated in Fig. \ref{figjp8}.
However, practical considerations mean that runs with large $Q'$ cannot be followed
to libration of the resonant angles. 
Nonetheless  when $Q'=100,$ $\cos(\Phi_3)$ and $\cos(\Phi_1)$
begin to attain non zero running means at the end of  simulations with minimum
masses after $2.5\times 10^7 yr.$
This indicates that resonance could be effectively entered for $Q' < 10^4$
within the lifetime of  such systems.
 These results  taken together suggest that the  planets may be undergoing
circularization in   a three planet resonance.

As the circularization process involves energy dissipation
at fixed angular momentum, the system must spread increasing the period ratios
of neighbouring planets. Although this means an increasing  departure
from strict commensurability, resonant angles remain librating
or circulating in such a way as to enable 
 the required redistribution of angular momentum to take place.
A simple model illustrating how this works is given in appendix A.
The amount of spreading and increase in the period ratios that can take place
in a given time depends on $Q'.$ We  investigate this
in the context of a migration scenario for producing the configuration below.

 Simulations   incorporating migration indicate that the  planets  were likely to have been in 
a strict Laplace resonance with all four angles librating.  
However, if this persists the period ratios  would increase
maintaining a larger departure from commensurability for the outer pair than the  inner pair
in disagreement with observations.
Thus some perturbation event must have disrupted this resonance
 enabling the planets to continue circularizing with the correct
period ratio relationships.  As we have seen above, giving
one of the planets a relatively small eccentricity 
can produce significant changes to the period ratios.

To investigate this scenario, we  now describe
simulations of systems of three~planets migrating inwards in a disc  that
set up  a Laplace  resonance.  We go on to study   the  evolution
 following from its disruption as the result of a
perturbation applied once the planets have decoupled  from the disc.  
\section{Migration and interaction with the star: Simulation results}\label{sec:simulations}

In all the runs we have performed, the planets converge on  each other 
forming a three  planet  4:2:1  resonance, and migrate in locked in that configuration. 
Whether
the resonance can be maintained all the way down to the disc inner
cavity depends on how fast the  migration is.  There is a tendency for the
4:2:1  resonance to be stable for slow migration rates but to be  disrupted, 
with the appearance of  higher order resonances 
 once the  migration  rate  becomes sufficiently  fast. 
 However, the evolution  is difficult to
predict precisely, as the transition from the 4:2:1  resonance to
higher order resonances happens through an instability which appears to
be extended in time and  chaotic. Small changes to the model parameters in this regime
may affect whether this instability occurs or not within a given time interval.
 Similar phenomena have been seen in two planet resonant migration
where a sequence of  resonances can be formed, maintained  and subsequently disrupted
in turn (see Papaloizou \& Szuszkiewicz 2009).

\begin{figure}
\includegraphics[width=16cm]{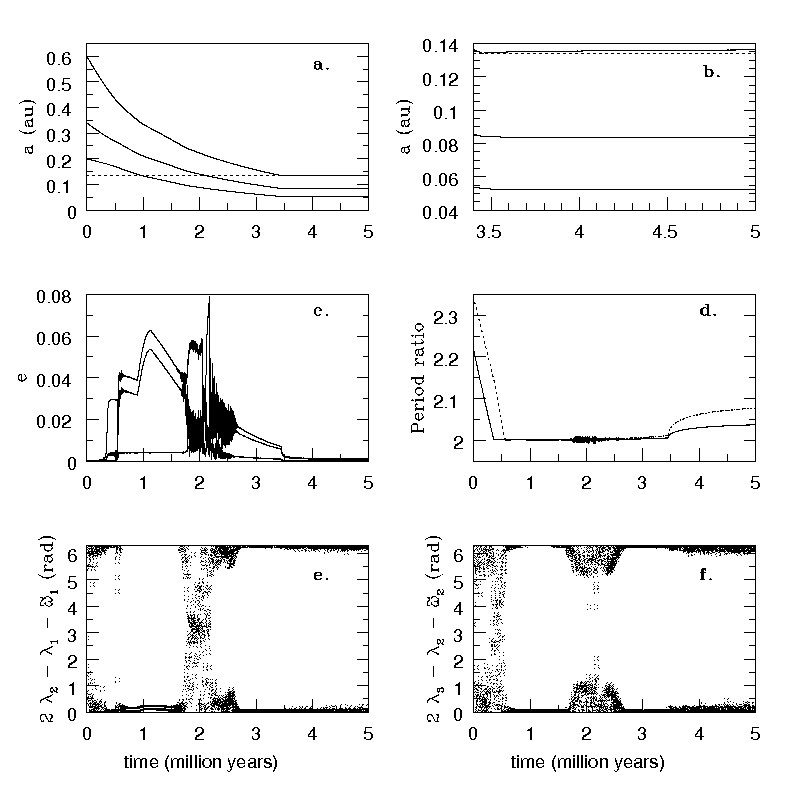} 
\caption{Semi--major axis $a$ (in au, {\em plots
 a} and {\em b}), eccentricity $e$ ({\em plot c}), period ratios ({\em
 plot d}), resonant angles $\Phi_{1}=2 \lambda_2 - \lambda_1 -
 \tilde{\omega}_1$ (in rad., {\em plot e}) and $\Phi_{3}=2 \lambda_3 -
 \lambda_2 - \tilde{\omega}_2$ (in rad., {\em plot f}) versus time (in
 million years) for 3 planets initially located at 0.2, 0.34 and 0.6 au.
 On plots a and b, each solid line represents a planet, whereas the
 dotted line indicates the location of the disc inner edge.
 Note the different time  interval  for  plot b.  On plot c, the solid and
 dotted lines represent the period ratio of the inner and outer pair of
 planets, respectively. Here $M_d=2.5 \times 10^{-5}$~M$_{\odot}$ and
 $Q'=10^{-2}$ (note that the evolution during the disc migration 
 phase is not sensitive to the value of this parameter).
  Even though strict commensurability is lost as a result
 of tidal interaction with the star, the planets stay locked in
 resonances, as indicated by the behaviour of the  resonant angles. }
\label{fig3}
\end{figure}

Figure~\ref{fig3} shows the evolution of the semi--major axes of the
planets and of the period ratios for 3 planets  with  minimum masses initially located at 0.2,
0.34 and 0.6~au, respectively. The disc aspect ratio is $H/r=0.05.$ 
 Equations (\ref{tm}) and (\ref{te})
for the migration and circularization times are used  with
 $M_d= 2.5\times10^{-5}$~M$_{\odot}.$ This was equal to the
disc mass within 5~au in the original formalism of Papaloizou \& Larwood (2000). 
However, as there is much uncertainty in the migration rates to be used
for low mass planets  we investigate different possibilities using  this parameter to scale the migration
and circularization rates (see section \ref{TypeImigration}).
Used in this way it no longer corresponds to a disc mass as in the original
formulation. 

The initial  migration times for this case  are
then  $2.8 \times 10^5$ and $4\times10^5$~years for the innermost and outermost planets
respectively.
 The initial period ratios for the inner and outer  pairs are  $2.21$  and
$2.34$, respectively.  
This calculation  led to the setting up of a Laplace resonance
that is  maintained during the subsequent  evolution of the system,
 although for a finite period of a time around $2 \times 10^6$~years a
temporary weak  and inconsequential
instability was present (see below).

In these  calculations , the tidal $Q'$ parameter was set to $10^{-2}$.
This is at least 3 orders or magnitude smaller than the values supposed
to apply in Earth--like planets, but this value was chosen to give
evolutionary timescales that could be reasonably  handled numerically  
(i.e. runs that could
be done within weeks on a single processor). Tests carried out below
with more realistic values of $Q'$  indicate that  the evolution during
the disc migration phase is not affected by this specification.
Furthermore,  when the planets are inside the cavity, the scaling
given by equation~(\ref{Hscal}) holds,
  implying that increasing $Q'$  corresponds  to scaling to a  longer evolutionary timescale.
With $Q'=10^{-2}$, the eccentricity damping timescale given by
equation~(\ref{teccs}) is $t_{e,i}^s =400$~years for a 1~M$_{\earth}$
planet at 0.05~au, inducing a timescale on which the semi--major axis
evolves  of about $2 \times 10^8$~years for an eccentricity of $10^{-3}$
(see section~\ref{sec:startide}).

Figure~\ref{fig3} shows the evolution of the semi--major axes, the
eccentricities, the period ratios and the resonant angles $\Phi_{1}=2
\lambda_2 - \lambda_1 - \tilde{\omega}_1$ and $\Phi_{3}=2 \lambda_3 -
\lambda_2 - \tilde{\omega}_2$.  
 As pointed out above, the Laplace resonance for which all four
resonant angles librate  is
established while the planets migrate in the disc, after about $5 \times
10^5$~years.  The resonant angles  $\Phi_{1}$  and $\Phi_3$ stay close to zero.
 Then, while some of the
planets are still in the disc, a weak  instability appears for a finite
period of time (after about $2 \times 10^6$~ years) and the resonance is
weakened with the angles driven to large amplitude libration
with an occasional circulation.  But the
instability eventually dies out and the resonance is recovered before
the planets decouple from the disc. 
Note that this instability is not
always seen in the runs we have performed.
 As indicated above it  seems to be associated
with an approach to a transition to a resonance of higher degree. This is indicated by the fact that
 when the migration
rate was speeded up by a factor of $4,$ the outer pair of planets underwent a transition
to a $3:2$ resonance while the inner pair maintained a $2:1$ commensurability at the point at
which the semi-major axes had attained the same configuration.

During the migration phase,
the eccentricities are pumped up to several hundredths.  Because of the
low value of $Q'$, tidal interaction with the star begins to affect the
evolution of the system slightly before the outermost planet penetrates
inside the cavity.  Once the outermost planet reaches the disc inner
edge, the disc is removed altogether from the calculation, so that if
the planet is pushed out again it  will not interact with the disc.
This procedure is introduced to take into account the
disappearance of the disc on a timescale of a few million years.  In
this particular run, the outermost planet enters the cavity after $3.5
\times 10^6$~years.  At that point, changes in semi--major axes are due
only to tidal interaction with the star.  As the eccentricities of the
two inner planets are damped from a few $10^{-3}$ to $10^{-3}$, their
semi--major axes decrease by about 1~percent (see plot b of
fig.~\ref{fig3}) in such a way that their period ratio increases by
  somewhat more than a percent, going from 2 to 2.037 (see plot d of
fig.~\ref{fig3}). 

 In the meantime, the eccentricity of the outermost
planet is hardly affected, as it was already very small (a few
$\times 10^{-4}$) when the planet entered the cavity.  Because of the
conservation of the total angular momentum though, its semi--major axis
increases by about 1~percent, which results in the period ratio of the
outer pair of planets increasing by a few percent, from 2 to 2.077.
Therefore, strict commensurability is lost.  However, we note that the
resonant angles still librate, so that the
Laplace resonance is not disrupted. The fact that the deviation of the outer period ratio
from~2 is twice the deviation of the inner period ratio from~2
is a consequence of the Laplace relation (equation~[\ref{LAPR}] of appendix A)
 and this feature is preserved
in the subsequent evolution of this and similar cases. As we shall see,
these deviations increase approximately $\propto t^{1/3}$, as predicted from equation~(\ref{Hscal}).

As noted above, the Laplace relation is not satisfied in HD~40307. As we shall see
below, it is possible to obtain the correct relationship between the period ratios
provided the Laplace resonance is disrupted such that only $\Phi_1, \Phi_3$ and $\Phi_4$
librate and contribute to the secular evolution (see also section \ref{initecc}) .

In order to investigate the dependence on the specification
of the circularization rates induced by the disc, we have performed
simulations for which these depend only on the planet mass and are therefore
constant.  We here consider  examples for which:

\begin{equation}
              t_{m,i}=7\times10^6 \;
             \frac{{\rm M}_{\oplus}}{m_i} \; \; {\rm yr},
 \end{equation}
and: \begin{equation}
        t_{e,i}^d=8\times10^3 \;
      \frac{{\rm M}_{\oplus}}{m_i}  \; \; {\rm yr}.
 \end{equation}

\begin{figure}
\begin{center}
\includegraphics[width=16cm]{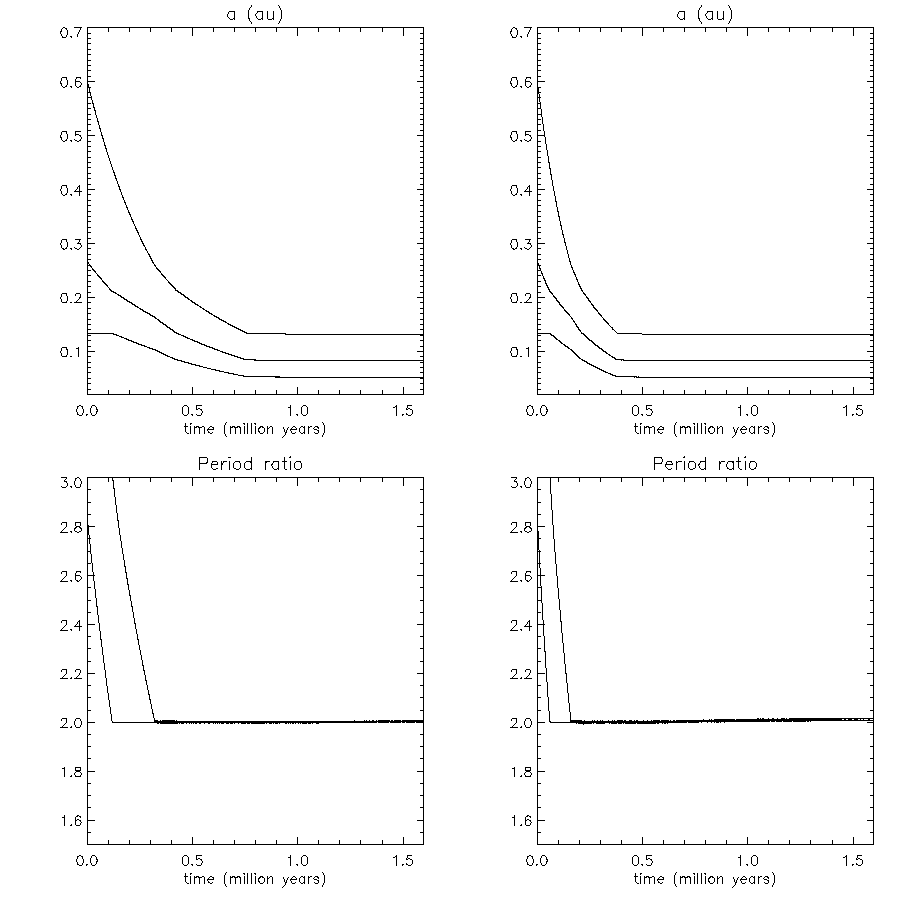} 
\caption{ The evolution of model
systems with the minimum masses of HD 40307 ({\em left panels})
and twice the minimum masses ({\em right panels}) showing the formation
and maintenance of three planet resonance as a result of disc migration.
For these models, the migration and circularization rates are constant and $Q'=10$
(although evolution during the disc migration phase is not sensitive to this parameter)}.
The {\em upper panels} give the evolution of the semi--major axes and the
{\em lower panels} the period ratios.
\label{figjpmig}
\end{center}
\end{figure}

Results for  cases  both  with the minimum masses for HD 40307 and
twice the mimimum masses are plotted in Figure \ref{figjpmig}.
During the disc induced  migration,
a three planet resonance is formed in both cases. As expected, the evolution
for the higher mass case is twice as fast but otherwise similar.
The resonance shows only  small signs of instability while the planets
migrate inwards. This is probably helped by the fact that the migration rate
is here constant rather than increasing inwards as in the previous simulation.
As the migration continues, the resonant angles $\Phi_i$ ($i=1,...,4$) all enter libration,
but after the planets enter the cavity there is an adjustment so that only
 $\Phi_{1}$ and $\Phi_3$ librate. The system then evolves back towards
a Laplace resonance where  all angles librate or contribute to long term time averages.
This latter aspect is similar to what was seen in the simulations with larger
$Q'$ illustrated in Figs.~\ref{figjp1}--\ref{figjp4}.

This final evolution is one where the eccentricities all  reduce
and the planets physically separate with increasing period ratios for the inner and
outer pairs of planets.
This form of evolution is to be expected in a Keplerian system in which energy
is dissipated while the total angular momentum
is conserved, as is the case for an accretion disc  (e.g., Lynden--Bell and Pringle 1974).
It is this form of evolution we investigate below and how far the tidal effects
of circularization can bring the system towards the HD~40307 system.
 
\subsection{Disruption of the resonance}

 As the continuation of the simulations described above
 maintains a three planet Laplace resonance, the period ratios will not have the same form
as in HD 40307. As indicated above the system has a larger period ratio  for the inner pair
as compared to the outer pair whereas the Laplace relation leads to the  converse.
However, we have found that there are resonant interactions for which only
$\Phi_1,\Phi_3,$ and $\Phi_4$ librate. Hereafter we refer to this as the standard state
of libration.  This  can lead to the  required form
for the period ratios but some disruption of the form of the resonance
initially set up within the cavity needs to take place.   Mechanisms
could involve close encounters or collisions  occurring   shortly
after the disc migration phase.

We comment that we  have found it impossible to disrupt the 2:1 resonance of the inner
pair of planets with a slow process.  For instance, we have added a
fourth planet on an inclined orbit, but found that even a massive
perturber (3~Jupiter masses) at 5~au could not take the inner pair away
from the 2:1 resonance.   To do this,
the  parameters  of one or more of the orbits
 have  to be varied suddenly  preventing the resonance from  responding adiabatically.
 An encounter  with a fourth planet on a parabolic orbit is a possible 
means to achieve this and there are other possibilities involving  one or more direct collisions.
To affect the resonance,
 the semi--major axis of one of the planets should
change by more than an effective libration width associated with 
one of the resonant angles. We have performed simulations for which the
velocity components of one of the planets  were all suddenly  increased or decreased
by a constant factor. This has the effect of preserving circular orbits 
which is reasonable under conditions of effective circularization.
We found that fractional changes exceeding $\sim 2\times 10^{-3}$  were sufficient
and, provided the changes were not too excessive, simulations approach the same form of evolution.

 Below we describe a simulation where an 
impulsive change  that  decreased the velocity components
of the innermost planet by $0.996$  was applied just after
the outermost planet entered the cavity.  
  The  end of the simulation illustrated in Fig.~\ref{figjpmig} 
  (or its larger planet  mass counterpart where appropriate)  was used as a  starting point
for the simulations presented below, as these are 
reasonable outcomes for a system having undergone
inward migration induced by a disc and then entered    an inner cavity.
 In some cases, the tidal parameter $Q'$  was also
adjusted.

 We remark that we have performed simulations with other types  of
applied impulse. For example we have augmented the above form
by giving the innermost planet an initial eccentricity in a similar manner
to that undertaken in section \ref{initecc}. When this is small and typically of order
$10^{-3}$ this makes no difference on account of  rapid effective circularization.
We have also considered impulsive changes to the outermost planet
  giving it an eccentricity $\sim 0.1$  as in section \ref{initecc}.
In this case period ratios can be shifted to be close to the observed ones
just after the outermost planet  has entered the cavity (see below).

\subsection{Evolution of systems on three planet resonances undergoing circularization
and radial spreading}\label{Results}

\begin{figure}
\begin{center}
\includegraphics[width=16cm]{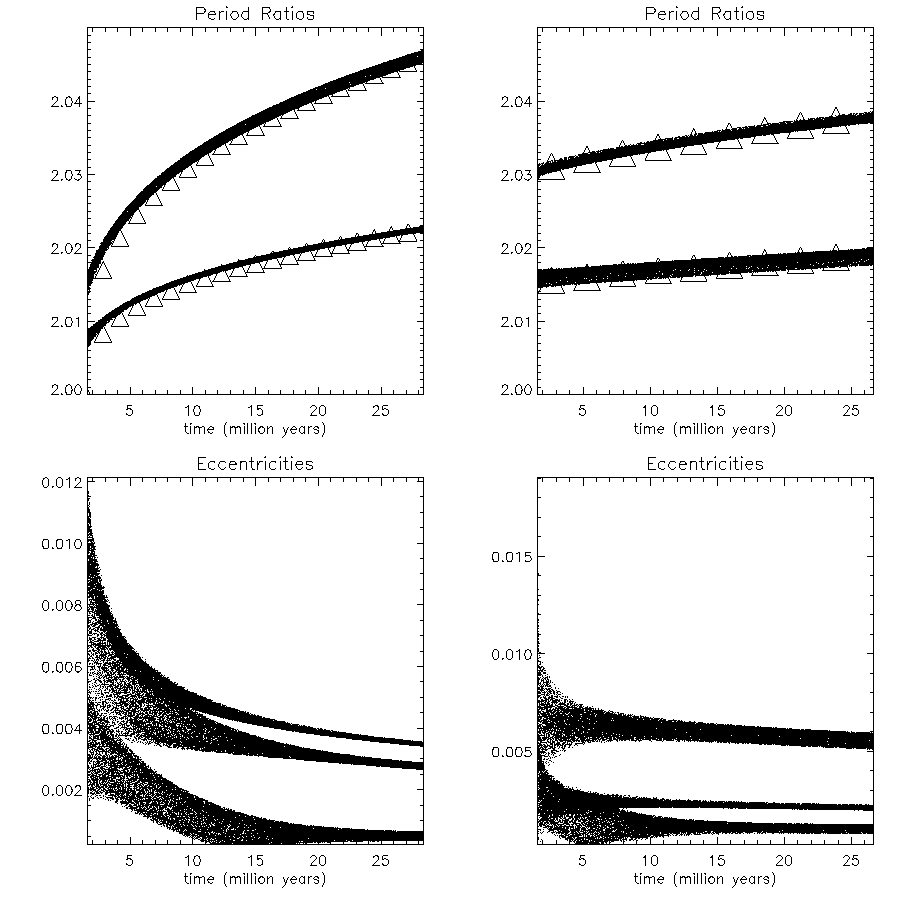} 
\caption{   Evolution of a model system of three
planets with masses twice the minimum masses of HD~40307.
The {\em left panels} correspond to a continuation of the evolution
from the corresponding case illustrated in Fig.~\ref{figjpmig}.
The {\em right panels} correspond to the evolution after an impulse was applied to the
inner planet (see text). For both models $Q' = 10.$
The period ratios $n_1/n_2$ and $n_2/n_3$ are plotted as a function
of time  in the {\em upper panels}. The upper curve corresponds to the outer pair
({\em left panel}) and the inner pair ({\em right panel}). In this and similar
figures below, the triangles correspond to analytic fits of the form
given by equation~\ref{anfit}) (and see text). 
The {\em lower panels} show the evolution of the eccentricities
of the three planets. In the {\em lower left panel} the curves from top down correspond to the inner,
middle and outer planets respectively. In  the {\em lower  right panel} the curves from top
down correspond to the middle, lower and outer planet.}
 \label{figjpmig1}
\end{center}
\end{figure}

In  Fig. \ref{figjpmig1}, we illustrate the evolution of  a model system of three
planets with masses twice the minimum masses of HD~40307 with $Q'=10.$
Two runs are displayed.  The first one continues  the evolution of the corresponding case
 plotted in~Fig.~\ref{figjpmig}.  In the second run,
an impulse was first applied to the innermost  planet that reduced
all velocities by a factor of $0.996.$ 

 As shown in Fig.  \ref{figjpmig1}, during the subsequent evolution the system spreads radially with increasing 
period ratios. The way this happens is  illustrated by the simple model
described in appendix A. 
In this discussion the period ratios are  replaced by the 
 quantities $W_{1,2} = n_1/n_2 -2 $ and $W_{2,3}= n_2/n_3 -2 $ 
 that are easily obtained from Fig.~\ref{figjpmig1}. 
We have  been able to find  good fits to the time evolution of the form 
 found in appendix A as  given by equation~(\ref{Hscal})
\begin{equation}
W_{i,i+1}=  \left( \frac{t_6-\beta}{\alpha} \right)^{1/3}, \; \; \; i=1,2,  \label{anfit}
\end{equation}
where $\alpha$ and $\beta$ are constants and $t_6$ is the time
in units of $10^6$~yr measured from the start of the relevant simulation
shown in Fig.~\ref{figjpmig}. 
 
 In the case with no initial impulse,
$\alpha= 2.77\times 10^5$ and $\beta=1.5 $
for the outer pair,  and $\alpha= 2.37\times 10^6$ and $\beta= $1.5 for the inner pair.
 For the case with an initial impulse,
 $\alpha= 7.73\times 10^6$  and  $\beta=-26.5 $
for the outer pair,  and $\alpha=9.66\times 10^5 $ and $\beta=-26.5 $ for the inner pair.
For the latter case, a standard libration state occurred with the deviation of the
inner period ratio from two being about a factor of two larger than the  corresponding
deviation for the outer period ratio, similar to the situation in HD~40307.
This is in contrast to the case with no impulse that persisted in a Laplace resonance
with all angles librating. In that case, the Laplace relation persists which leads
to the relative deviation of the period ratios from two behaving
differently from HD~40307 (see above).

Despite this, we can estimate the time it would take
to obtain the outer period ratio of 2.13 appropriate to HD~40307. This is
$6.1\times 10^8$~yr. The scaling of the circularization time,
$t_{e,i}^s \propto m_i^{-2/3}$ leading to the scaling
of the evolution time, $t,$  to reach a given state,  with mass of the form   $t \propto m_i^{-8/3},$
 implies that for a minimum mass system this would be $\sim 4\times 10^9$~yr.
Thus, it might just be possible to achieve the outer period ratio in this case but then
the inner period ratio would have to be determined subsequently by alternative  means.

For the case with the applied  impulse, we may
estimate the time to bring the inner period ratio
to the appropriate value $2.23.$ This is somewhat uncertain because
of the small variation seen in the simulation, but it is about  $\sim 1.1\times 10^{10}$~yr.
Note that the relative period ratios are not precisely identical to those
of HD~40307.  In particular the ratio of the deviation of the inner  period
ratio from two and,  the deviation of the outer period ratio from two  is about $30\%$ too large. 
However,  this may be adjusted by changing  simulation  input  parameters.
For  example we recall that the simulations reported here
 were performed with the same value of $Q'$
for all of the planets.  To indicate one possibility we note that
 tests we have performed indicate that  the above ratio
may be reduced  by taking a smaller  value of $Q'$ for the outermost planet
than the others.

We remark that, while it might be barely  possible to obtain an inner period ratio of 2.23 for a
larger mass system,  it is almost certainly not possible for the minimum mass system
as the time required would be $\sim 6.98\times 10^{10}$~yr.

Note that, as the evolution progresses, the eccentricities of all the planets tend to decrease,
as does their variation. This is a general  consequence of the spreading that is driven by
the energy dissipation in these simulations.

\begin{figure}
\begin{center}
\includegraphics[width=16cm]{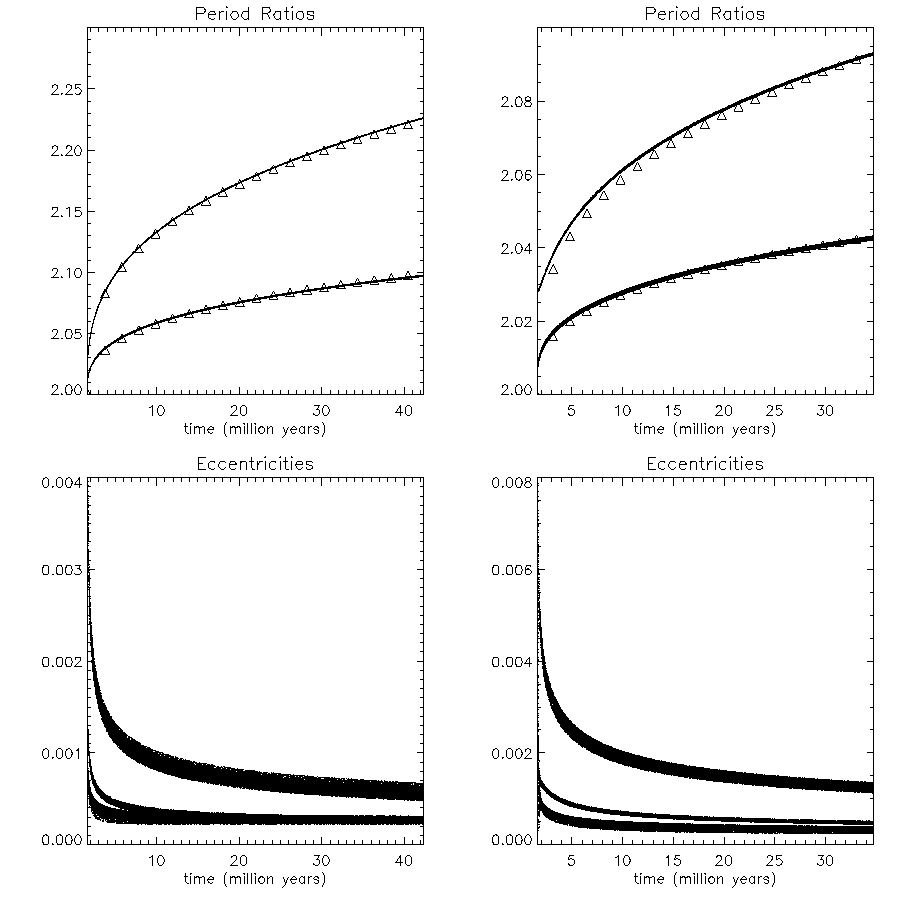}
 \caption{   As in Fig.~\ref{figjpmig1},
but for models with minimum masses and for which an impulse
was applied to the innermost planet (see text).
For the {\em left panels}, $Q'=0.01$ and for the {\em right panels}, $Q'=0.1.$
In the {\em upper panels},  the upper curves correspond to
the inner period ratio.
In the {\em lower panels}, the curves from top down correspond to the
middle, inner and outer planets respectively.}
 \label{figjpmig2}
\end{center}
\end{figure}

To investigate further the form of the evolution and its scaling
with run parameters, we plot in Fig.~\ref{figjpmig2} the results of two simulations
 of  models with minimum masses and for which an impulse
of the above form was applied to the innermost planet. 
We compare cases with
 $Q'=0.01$ and  $Q'=0.1.$ As stated above, we had to consider rather small values in order to perform
simulations that produced a reasonable variation in a reasonable time.
These resulted in a standard state of libration and relative period ratios very similar
to the higher mass simulation with an initial impulse illustrated in Fig.~\ref{figjpmig1}.

\noindent In the case with $Q'=0.01$, the fit parameters are
$\alpha= 3.58\times 10^3$ and  $\beta=1.7 $
for the inner pair,  and $\alpha= 4.20\times 10^4$ and $\beta= $1.7 for the outer pair.
For the case with $Q'=0.1$, they are
 $\alpha= 4.13\times 10^4 $ and $\beta= 1.5.$ for the inner pair,  and $\alpha= 4.18\times 10^5$ 
and $\beta= $1.5 for the outer pair. The values of $\alpha$, being about an order of magnitude larger
in the second case, are consistent with the evolution time being an order of magnitude
larger, thus scaling with $Q'$ as expected.
The time to attain an inner period ratio of $2.23$ in the $Q'=0.01$ case
is $\sim 4.4\times 10^7$~yr,  leading to an estimate of $\sim 4.4\times 10^{10}$~yr
when $Q'=10.$  This is less than the estimate based on the analytic scaling  with planet mass
implied by
equation (\ref{Hscal}) applied to
the higher mass simulations illustrated in Fig.~\ref{figjpmig1} by a factor $\sim 1.5,$
which we  believe is a reasonable  level of consistency 
in view of the extent of the extrapolation involved.
The eccentricities  in these simulations 
decrease with time in a manner analogous to
the simulations shown in Fig.~\ref{figjpmig1}.

 The long times required for the above cases to reach a configuration similar
to that now observed in   HD~40307 arise   because only a very small perturbation
was applied  just after the systems decoupled from the disc, requiring a large amount
of subsequent period ratio  evolution. Such a scenario is only feasible for
masses exceeding twice the minimum and small $Q' \sim 10.$

However, if a larger perturbation is allowed, it can provide a larger shift in 
the period ratios requiring less evolution later on (see Fig. \ref{figjp7}).
 This is equivalent
to starting on the evolutionary curves shown in Figs. \ref{figjpmig1}-\ref{figjpmig2} at a later time.

For example we found that if a perturbation of the type adopted in section~\ref{initecc}
was applied to the outer planet a shift of the  right type  could be obtained.
We found that when the imposed  eccentricity of the outer planet 
was taken to be $0.15,$ the period ratios were shifted to very similar
 values to those in HD~40307.
Thus if large enough perturbations are allowed, only a small amount of subsequent
evolution may be needed.

\section{Discussion and conclusion}
\label{sec:discussion}

In this paper, we have shown that the planets around the star HD~40307 could be undergoing a resonant interaction, despite departure 
of the period ratios from very precise commensurability.   This is indicated by the fact that  three of the four
 resonant angles librate
or are associated with long term changes to the orbital elements.   Note that such a resonant state can occur without exact commensurability only because the eccentricities are very small, which can be brought about as a result of tidal circularization  from a state with initially higher eccentricities
(see section \ref{initecc}).
When this occurs in a system that starts from a near 2:1 commensurability, the difference of the period ratios from two
then increases $\propto t^{1/3}.$

We propose that the planets in this system were in a strict Laplace resonance while they migrated through the disc, with all 4 resonant angles librating.   Exact commensurability was  then departed from as indicated above as a result of  the tidal interaction with the star, which preserved libration of at least some of the resonant angles, after the planets entered the disc inner cavity.  Because of the Laplace relation ,
  the period ratios evolve in such a way that the ratio for the inner pair of planets is smaller than that of the outer pair of planets.  The opposite is observed in HD~40307. 
  
   To get the appropriate form for the period ratios, some  disruption of the resonance set up in the cavity needs to take place (this is in contrast to claims by Zhou 2009).  A  close encounter for instance 
 might be a possible cause.   An impulsive interaction would be sudden enough that the resonance would not  respond adiabatically.   If the velocity of the innermost planet is reduced by a factor 0.996 for instance,  deviation of the inner period ratio from two gets twice as large as the corresponding deviation for the outer period ratio, while 3 of the 4 resonant angles still librate,  similar to the situation in HD~40307,  within the lifetime of the system. 
 
  However, we found that  we could get a period ratio of 2.23 for the inner pair  only if the masses are significantly larger than the minimum masses and tidal dissipation is possibly unrealistically effective. 
 From the results obtained in section \ref{sec:simulations}, when the Laplace resonance is disrupted at an early stage,
   we estimate that the deviation of the inner period ratio from two
  is approximately given as a function of time by    $W_{1,2} \sim  0.23( 0.157 \tilde m_1^{8/3} t_{10} /Q'_{10})^{1/3},$
  where $Q'_{10}=Q'/10,$ $t_{10} = t/(10^{10}~{\rm yr})$  and $\tilde m_1 =  m_1/  4.2~{\rm M}_{\oplus} $ ($m_2/m_1$ and $m_3/m_1$
  are assumed fixed for the purposes of this discussion).
  Accordingly, to attain  $W_{12} =0.23,$ we would require  masses exceeding the minimum
  estimate by about a factor of two for $Q'_{10} \sim 1.$
  
   If  the Laplace resonance is preserved, 
  from the results obtained in section \ref{sec:simulations}, we may also estimate that the deviation of the outer period ratio from two
  is approximately given as a function of time by    $W_{1,2} \sim  ( 2.6 \tilde m_1^{8/3} t _{10}/Q'_{10})^{1/3}$.
  Thus it may  be  possible to get a period ratio of 2.13 for the outer pair of planets 
  with near to minimum masses  in a reasonable time if  $Q'_{10}\sim 1$  for all the planets,
   but then the inner period ratio would be only $\sim 2.065.$ 
   It would thus seem likely that additional  stronger 
   dynamical interactions are required to bring the system to
   its final state.

   For example we found that if the outer planet was given an eccentricity $e_3\sim0.15$
   shortly after decoupling from the disc, period ratios similar to those observed in HD~40307
     could be  produced  on a time scale of  a few~$\times10^6 yr.$
   (see sections \ref{initecc} and \ref{sec:simulations}). 
   Then, the work presented in this paper indicates that tidal circularization would
   continue through resonant interaction with the period ratios separating after these interactions are over. 

Note that  a similar type of scenario may apply to the system of planets around GJ~581, which is currently closer to exact commensurablilty than HD~40307.  Indeed, different mean motion resonances can be established while the planets migrate within the disc, depending on the initial relative location of the planets and on the migration timescale.  However, as it appears that this system
may have been in higher order resonances than HD~40307,
 it is possible that these would have been broken completely as a result of circularization.
  We intend to present a study of  GJ~581 in a forthcoming paper.  

In a study of the tidal evolution of the system around HD~40307, Barnes et al. (2009) have concluded that the planets cannot be Earth--like.   According to them, if it were the case, extrapolating back in time the tidal evolution of the  eccentricities from  current values
 (estimated from direct simulations of the  observed system),  the system would have been unstable.   The work in this paper does not support such a conclusion.  We note that Barnes et al. (2009) neglected the dynamical  interactions between the planets,  which clearly will have  played a role 
if the  current eccentricities are small because   resonant interactions  coupled with tidal effects have been associated 
with long term changes to the orbital elements.  

The results reported in this paper indicate  that resonances in multiple low mass--planet systems may 
play more of a role  than  currently claimed in the literature.  Significant departure from exact mean motion resonance does not 
necessarily  preclude a resonant state when, for example, tidal effects are important causing eccentricities to be small, 
as libration of some of the resonant angles in that case may still exist,  and be  associated with long term evolution of the orbital elements.    Although  the  confirmation of resonance effects is problematic for small eccentricities, we expect that  further detections  and analysis of systems similar to HD~40307 will  test the relevance of
the scenarios explored in this paper.  

Finally, we comment that, because of the rapid increase of the effectiveness of tidal interactions 
with decreasing orbital period,
 the dynamical interactions discussed here would be of much  greater importance 
for systems in which the period of the innermost planet is significantly
 shorter than in the HD~40307 system.  For example, we note that
 the planet CoRoT--7~b, which has a projected mass of 11.12~M$_{\oplus}$, has 
a period of only 0.85 day (L\'eger et al. 2009).   If this short period orbit is a result of migration,
 it is likely this planet was pushed inwards by a companion
which would have been on a commensurable orbit (Terquem \& Papaloizou 2007).
 We note in support of this idea 
that pre-main sequence stars typically  have rotation periods of several days (Bouvier et al. 1993), which would
imply that the disc inner cavity should extend well beyond
 the orbit of  CoRoT--7~b if that  were magnetically maintained. In that case, 
this planet  could not have migrated so far inwards without being shepherded by a companion.

\begin{appendix}
\section{Semi--analytic model for a system in a three planet resonance 
undergoing circularization}
\label{sec:analysis}

In this appendix we  develop a semi--analytic model that shows how a two or three
planet system undergoes radial spreading as a result of evolution driven by tidal circularization.
The coupling between the planets occurs because resonant angles may librate
or be associated with long term changes {\it even though there may be significant
deviations from exact commensurability}.  
We begin with the system
without tides which is Hamiltonian.  Subsequently, we add the effects of tides.

\subsection{Coordinate system}
We adopt Jacobi coordinates (Sinclair~1975, Papaloizou \& Szuszkiewicz~2005) for which the radius vector
 of  planet $i,$  ${\bf r }_i,$ is measured relative
to the  centre of mass of  the system comprised of $M$ and  all other  planets
interior to  $i,$ for  $i=1,2...,N$ with $N=3$ here and $i=1$ corresponding to the
innermost planet. 
The  Hamiltonian,  correct to second order in the planetary masses,  can be written in the form:
\begin{eqnarray} H & = &  \sum_{i=1}^3 \left({1\over 2}  m_i | \dot {\bf r}_i|^2
- {GM_{i}m_i\over  | {\bf r}_i|} \right)   \nonumber \\
& - &\sum_{i=1}^3\sum_{j=i+1}^3Gm_{i}m_j
\left({1 \over  | {\bf r}_{ij}|}  -  { {\bf r}_i\cdot {\bf r}_j
\over  | {\bf r}_{j}|^3}\right).
\end{eqnarray}
Here $M_{i}=M+m_i $ and
$ {\bf r}_{ij}= {\bf r}_{i}- {\bf r}_{j}.$

The equations of motion for motion  
in a fixed plane, about a dominant central mass,
 may be written in the form
(see, e.g., Papaloizou~2003, Papaloizou \& Szuszkiewicz~2005):

\begin{eqnarray}
\dot E_i &=& -n_i\frac{\partial H}{\partial \lambda_i}\label{eqnmo1}\\
\dot L_i &=& -\left(\frac{\partial H}{\partial \lambda_i}+\frac{\partial H}{\partial \varpi_i}\right)\\
\dot \lambda_i &=& \frac{\partial H}{\partial L_i} + n_i \frac{\partial H}{\partial E_i}\\
\dot \varpi_i &=& \frac{\partial H}{\partial L_i}.\label{eqnmo4}
\end{eqnarray}
Here the orbital  angular momentum of  planet  $i$ which has
reduced mass
$m_i = m_{i0}M/(M+m_{i0}),$ with $m_{i0}$ being the actual mass,  is $L_i$ and the
orbital  energy is $E_i.$
For motion around a central point  mass $M$ we have:
\begin{eqnarray}
     L_i &=&  m_{i}\sqrt{GM_{i}a_i(1-e_i^2)}, \\
     E_i &=& -{{GM_{i}m_{i}}\over{2a_i}},
\end{eqnarray}
where $M_{i} = M+m_{i0},$  $a_i$ denotes the semi-major axis and $e_i$  the eccentricity
of planet $i.$

\noindent The  mean longitude of planet $i$ is $\lambda_i = n_i (t-t_{0i}) +\varpi_i ,$
 where $n_i  = \sqrt{GM_{i}/a_i^3}$ is its mean motion, with
$t_{0i}$ denoting its time of periastron passage 
and $\varpi_i$ the longitude of periastron.
We   remark that results valid for a two planet system, $i =1,2$, may  be obtained
by setting $m_3 =0.$

\noindent The Hamiltonian may quite generally
 be expanded in a Fourier series
involving linear combinations of the  five angular differences
$\varpi_1 -\varpi_2,$ $\varpi_2 -\varpi_3$ and
$\lambda_i - \varpi_i, i=1,2,3. $ 

 Here we are interested
in the effects of two first order  $2:1$ commensurabilities
associated with the outer and inner pairs of planets.
In this situation,  we expect that any of the four angles
$\Phi_1 = 2\lambda_2-\lambda_1-\varpi_1, $ 
$\Phi_2 = 2\lambda_2-\lambda_1-\varpi_2,$
$\Phi_3 = 2\lambda_3-\lambda_2-\varpi_2 $ and  
$\Phi_4 = 2\lambda_3-\lambda_2-\varpi_3$
may  be slowly varying.
 Following  standard practice
 (see, e.g., Papaloizou~2003, Papaloizou \& Szuszkiewicz~2005),
only terms in the Fourier expansion involving  linear
combinations of $\Phi_1,$  $\Phi_2,$ $\Phi_3$ and $\Phi_4$
as argument are  retained. 
Working in the limit of small eccentricities applicable here,
only terms that are first order in the eccentricities need to be retained.

\noindent 
Expanding to first order in the eccentricities, the  Hamiltonian   may  be written
in the form:
\begin{equation} H=E_1+E_2+ E_3 +  H _{12} +H_{23},\label{Hamil0} \end{equation}
 where:
\begin{equation} H _{ij}= -\frac{Gm_im_j}{a_j}\left[  e_j C_{i,j}\cos (\Phi_{i+j-1})
- e_iD_{i,j}\cos (\Phi_{2i-1}) \right], \label{Hamil} \end{equation}
with:
\begin{eqnarray} C _{i,j} & = & {1 \over 2}\left(   x_{i,j}{db^{1}_{1/2}(x)\over dx} \Biggl|_{x=x_{i,j}}\Biggr. +3b^{1}_{1/2}(x_{i,j})
-4x_{i,j} \right),   \label{Hamil1} \\
D _{i,j} & =& {1 \over 2}\left(   x_{i,j}{db^{2}_{1/2}(x)\over dx}\Biggl|_{x=x_{i,j}}\Biggr. +4b^{2}_{1/2}(x_{i,j})  \right) . \label{Hamil2} 
\end{eqnarray}
Here $b^{p}_{1/2}(x)$ denotes the usual Laplace coefficient
(e.g., Brouwer \& Clemence 1961)
with the argument $x_{i,j} = a_i/a_j.$ 

\subsection{Behaviour near a resonance with disc tides incorporated}
 
 Using equations~(\ref{eqnmo1})--(\ref{eqnmo4})
together with equation~(\ref{Hamil0})  we may first  obtain the equations of motion
without the effect of circularization tides  and then add this in. We obtain 
correct to lowest  order in the eccentricities:
\begin{align}
\frac{d e_1}{dt}&=
\frac{m_2a_1 n_1D_{12}}{a_2 M_1}\sin\Phi_1 - \frac{e_1}{t_{e,1}^s}, \label{eqne1}\\
\!\!\!\!\!\!\!\!\!\!\!
\frac{d e_2}{dt}&=
-\frac{n_2 }{M_2}
\left(m_1C_{12}\sin\Phi_2-\frac{a_2}{a_3}m_3D_{23}\sin\Phi_3\right)
-\frac{e_2}{t_{e,2}^s},
\label{eqne2}\\
\!\!\!\!\!\!\!\!\!\!
\frac{d e_3}{dt} &=
-\frac{m_2 n_3C_{23}}{ M_3}\sin\Phi_4 - \frac{e_3}{t_{e,3}^s}, \label{eqne3}\\
\!\!\!\!\!\!\!\!\!\!
\dot n_1 &=
-\frac{3n_1^2m_2a_1}{M_1a_2}\left(C_{12}e_2\sin\Phi_2-D_{12}e_1\sin\Phi_1 \right)
+\frac{3n_1e_1^2}{t_{e,1}^s}, \label{eqntid0}\\
\!\!\!\!\!\!\!\!\!\!
 \dot n_2 &=
\frac{6n_2^2m_1}{M_2}\left(C_{12}e_2\sin\Phi_2-D_{12}e_1\sin\Phi_1 \right)\nonumber\\
\!\!\!\!\!\!\!\!\!\!
   &-
\frac{3n_2^2m_3a_2}{M_2a_3}\left(C_{23}e_3\sin\Phi_4-D_{23}e_2\sin\Phi_3 \right)
+\frac{3n_2e_2^2}{t_{e,2}^s}, \label{eqntid} \\
\dot n_3 &=
\frac{6n_3^2m_2}{M_3}\left(C_{23}e_3\sin\Phi_4-D_{23}e_2\sin\Phi_3 \right)
+\frac{3n_3e_3^2}{t_{e,3}^s},  \label{eqntid1}\\
\!\!\!\!\!\!\!\!\!\!
\dot \Phi_1 &=
2n_2-n_1-\frac{1}{\sqrt{GM_1a_1}m_1e_1}\frac {\partial H_{12}}{\partial e_1}\Biggl|_{a_1,a_2,a_3}\Biggr. ,
 \label{eqntid2}\\
\!\!\!\!\!\!\!\!\!\!
\dot \Phi_2 &=
2n_2-n_1-\frac{1}{\sqrt{GM_2a_2}m_2e_2}\frac {\partial (H_{12}+H_{23})}{\partial e_2}\Biggl|_{a_1,a_2,a_3}\Biggr. ,
 \label{eqntid3}\\
\!\!\!\!\!\!\!\!\!\!
\dot \Phi_3 &= 
2n_3-n_2-\frac{1}{\sqrt{GM_2a_2}m_2e_2}\frac {\partial (H_{12}+H_{23})}{\partial e_2}\Biggl|_{a_1,a_2,a_3}\Biggr. ,
 \label{eqntid4}\\
\!\!\!\!\!\!\!\!\!\!
\dot \Phi_4 &=
2n_3-n_2-\frac{1}{\sqrt{GM_3a_3}m_3e_3}\frac {\partial H_{23}}{\partial e_3}\Biggl|_{a_1,a_2,a_3}\Biggr. .
 \label{eqntid5}\\
\nonumber
\end{align}
The
 terms on the right hand sides involving   the circularization times
$t_{e,i}^s$  describe the  effects  arising from  tides
 raised by the central mass on  planet $i.$
The terms $\propto e_i^2/t_{e,i}^s$
in equations~(\ref{eqntid0}), (\ref{eqntid1}) and (\ref{eqntid2})
 account for the orbital energy dissipation occurring as a result of circularization
at the lowest order in $e_i$  for which it appears.

\subsection{Energy and angular momentum conservation}

In the absence of tides, the total energy $E\equiv H$ and angular momentum
$L=L_1+L_2+L_3$ are conserved.
When circularizing tides act, the total angular momentum is conserved but energy is lost according to:
\begin{equation}
\frac{dE}{dt}= \sum_{i=1}^3\frac{2e_i^2E_i}{t_{e,i}^s}.
\label{Econ}
\end{equation}

\subsection{Resonance  tides and  migration}\label{Analytic}
When  both pairs of planets $(1,2)$ and $(2,3)$ are in resonance, a three planet resonance
is said to exist.  It is convenient to perform  a time average
of equations (\ref{eqne1})-(\ref{eqntid5}) 
taken over a time long compared to both the characteristic orbital period
and the libration period ,  but short enough
compared to the  tidal time scale (see eg. Sinclair 1975 for a discussion of this aspect). 
 The angles $\Phi_i$  may be undergoing libration or circulation.

 Equations (\ref{eqne1})--(\ref{eqntid1}) may be averaged directly,
while equations~(\ref{eqntid2})--(\ref{eqntid5})
may be averaged directly for small amplitude librations or alternatively after multiplying them by
 quantities such as  either, e.g.,  $\cos\Phi_1,$ $\cos\Phi_2,$ 
$\cos\Phi_3$  and $\cos\Phi_4$   or $e_1\cos\Phi_1,$ $e_2\cos\Phi_2,$
$e_2\cos\Phi_3$  and $e_3\cos\Phi_4$, respectively.  The time averages of these
quantities  as well as  $\sin\Phi_i$ and $e_i\sin\Phi_i,$ $(i = 1,2,3)$
are assumed to vary on the tidal time scale or longer.
When these are non zero,
 long term changes to the orbital elements  $a_i$ and  $e_i$ may occur through
 planet-planet interactions even without tides.
As illustrated in Figures  \ref{figjp2} and \ref {figjp2a},  six of
 these  quantities are found to exhibit  steady long term time averages of the required type that are quickly established  for a model of HD~40307 with assumed $Q' = 1000.$

 As this averaging procedure results in equations of essentially the same form as the
 case where the participating resonant angles (defined to be those that are associated
 with non zero time averages)
 are assumed to have averages close to  but not  equal to zero or $\pi,$ 
 while undergoing  small amplitude librations,  leading to the same scaling with physical parameters, but with an adjustment of numerical coefficients.  For simplicity we  restrict consideration to that  case.
Assuming small  amplitude librations of all four angles, equations~(\ref{eqntid2})--(\ref{eqntid5}) give relations of the form:
\begin{eqnarray}
2n_2-n_1 & = & \frac{m_2n_1a_1D_{12}\cos\Phi_1}{e_1M_1 a_2},\label{lap1} \\
2n_2-n_1 & = & \beta, \label{lap2} \\
2n_3-n_2 & = & \beta, \label{lap3} \\
2n_3-n_2 & = & -\frac{m_2n_3C_{23}\cos\Phi_4}{e_3M_3 },\label{lap4}
\end{eqnarray}
where:
\begin{equation}\beta= \frac{m_3n_2a_2D_{23}\cos\Phi_3}{e_2a_3 M_2 }-\frac{m_1n_2C_{12}\cos\Phi_2}{e_2M_2 }
\end{equation}
and the $\Phi_i$ in the above can be set to  either zero or $\pi.$

 We note that, when all four angles
participate,  equations~(\ref{lap2}) and~(\ref{lap3})
imply the strict Laplace resonance relation 
\begin{equation} 3n_2-2n_3-n_1=0.\label{LAPR}\end{equation}
However,  simulations of HD~40307  indicate  that only $\Phi_1, \Phi_3, $ and $\Phi_4$
participate. Thus equation (\ref{lap2})
is absent and the strict  Laplace relation does not hold. Indeed,
this relation implies that  ${\cal L} =(n_1/n_2-2)/(n_2/n_3-2)n_2/n_3$
should be unity. Being about $4$,  the Laplace relation is not satisfied.
When $\Phi_2$ does not  participate and the Laplace relation does not hold,
the  quantity ${\cal L}$ is not constrained from the outset but
is determined as a consequence of the initial conditions 
and following time dependent evolution.

 Below, we consider
the case when either all four angles contribute or only $\Phi_1,\Phi_3$ and $\Phi_4$ contribute.
We note that a  similar discussion applies  when only  $\Phi_1,\Phi_2$ and $\Phi_3$ 
participate that leads to the same conclusions ($e_3$ decays independently
due to tides in that  case and may be taken to be zero).
In all cases,  one can readily verify that the above system provides a complete
set of equations for the  time averages of the 
resonant angles and the  orbital elements when the form of the tidal
forces is specified.

However, we note that we can obtain 
the characteristic form and time scale of the  evolution by considering
the conservation  energy 
given by  (\ref{Econ}) together with the constancy of the total angular momentum.
 To do this we begin, by noting  that
equations (\ref{lap1}) and (\ref{lap4}) provide equations for the mean motion ratios of the form:
\begin{equation}
 \frac{n_{2}}{n_1}= f_{2,1}(\xi_{2,1})  \ \ \ \
{\rm where} \ \ \ \
 \xi_{2,1} = \frac{m_2\cos\Phi_1}{e_1M_1}, \end{equation}
and:
\begin{equation}
 \frac{n_{3}}{n_2}= f_{3,2}(\xi_{2,1})   \ \ \ \
{\rm where}    \ \ \ \
 \xi_{3,2} = \frac{m_2\cos\Phi_4}{e_3M_3}, \end{equation}
and $f_{2,1}$ and $f_{3,2}$ are defined implicitly through equations~(\ref{lap1}) and (\ref{lap4}).
In addition, equation~(\ref{lap3}) gives  $e_2$ in terms of  $e_3$, $\xi_{2,1}$
and $\xi_{3,2}$ in the form:
\begin{equation}
\frac{e_2}{e_3} =  \frac{m_3M_3n_2a_2D_{2,3}\cos\Phi_3-m_1M_3n_2a_3C_{1,2}\cos\Phi_2}
{m_2M_2n_3a_3C_{2,3}\cos\Phi_4}.\label{e3eq}
\end{equation}
When $\cos\Phi_2$ has zero time average and so is  replaced by zero, 
equation~(\ref{lap2}) does not hold and accordingly the Laplace relation also does not hold. 
We comment that $\xi_{2,1}$ and $\xi_{3,2}$ measure the deviation
from strict commensurability of the corresponding mean motions
and {\it do not  have to be small.}  For a system such as HD~40307 these  are   large,
particularly in comparison  
comparison to the characteristic size of the eccentricities. 

We remark that according  to equations (\ref{lap1}) - (\ref{lap4}), other things 
remaining  fixed,  the eccentricities
are proportional to the planet mass scale and,    inversely proportional to
the deviation from commensurability. The results presented in section \ref{initecc}
confirm these trends but with some deviation that is to be  expected
because all the relevant resonant angles, although contributing to the 
evolution, are not in a state of small amplitude libration in those simulations.

\subsection{Energy and Angular momentum conservation}
 Replacing $M_i$ ($i=1,2,3$)
by $M$ to simplify expressions without loss of content (as $M \gg m_i$),  the total energy
and angular momentum may  be written, respectively,  as:
\begin{equation}
E=-\frac{GMm_1}{2a_1}\left[ 1+ \frac{m_2f_{2,1}^{2/3}}{m_1}+
\frac{m_3(f_{3,2} f_{2,1})^{2/3}}{m_1} \right],
\end{equation}
and:
\begin{equation}
L=m_1\sqrt{GM a_1} \left[ 1+ \frac{m_2}{m_1 f_{2,1}^{1/3}}
+\frac{m_3}{m_1(f_{3,2} f_{2,1})^{1/3}} \right].
\end{equation}
Here, we have neglected in $L$ terms proportional to the squares of the eccentricities
as these can be shown to lead to small effects compared to those arising
from the eccentricity dependence of $f_{2,1}$ and $f_{3,2}.$
We now use  the above together with equation~(\ref{Econ})  to form
the quantity $L^{-2}d(EL^2)/dt = dE/dt+(2E/L)dL/dt$ and find  that this is given by:

\begin{equation}
-\frac{GMm_2}{3a_2}\left(1
+\frac{2E}{Ln_2}\right)\frac{{\dot f_{2,1}}}{f_{2,1}}
-\frac{GMm_3}{3a_3}\left(1
+\frac{2E}{Ln_3}\right)\left( \frac{{\dot f_{2,1}}}{f_{2,1}}+ \frac{{\dot f_{3,2}}}{f_{3,2}}\right)
=
\sum_{i=1}^3\frac{2e_i^2E_i}{t_{e,i}^s}. \label{ELcon}\end{equation}

We may write equation~(\ref{ELcon}) in terms of the  deviations from commensurabiity:
$W_{2,1} =  n_1/n_2-2=f_{2,1}^{-1}-2$ and
 $W_{2,3}=n_2/n_3-2=f_{3,2}^{-1}-2$, obtaining:
\begin{equation}
\frac{GMm_2n_2}{3a_2n_1}\left(1
+\frac{2E}{Ln_2}\right){\dot W_{2,1}}
+\frac{GMm_3}{3a_3}\left(1
+\frac{2E}{Ln_3}\right)\left( \frac{n_2}{n_1}{\dot W_{2,1}}+\frac{n_3}{n_2} {\dot W_{2,3}}\right)
=
\sum_{i=1}^3\frac{2e_i^2E_i}{t_{e,i}^s}.\label{ELcon1}\end{equation}
In addition, the right hand side can be expressed in terms of $W_{1,2}$ and
$W_{2,3}$ using equations~(\ref{lap1})--(\ref{lap4}) which give:
\begin{equation}
\sum_{i=1}^3\frac{2e_i^2E_i}{t_{e,i}^s} =
\frac{Gm_1m_2^2n_1^2a_1D_{1.2}^2\cos^2\Phi_1}{W_{2,1}^2Mn_2^2a_2^2t_{e,1}^s}
+\left[1+ \left( \frac{e_2}{e_3} \right)^2 \frac{E_2t_{e,3}^s}{E_3t_{e,2}^s} \right]
\frac{Gm_3m_2^2n_3^2C_{2,3}^2\cos^2\Phi_4} {W_{2,3}^2Ma_3n_3^2 t_{e,3}^s}.
 \label{ELWcon}\end{equation}
We may now obtain characteristic form of  the time  evolution
of the departure from strict commensurability. To do this, we may simply assume that
$W_{2,1},$ $W_{2,3}$  are comparable, significantly less than unity
  and $\sim W_0,$ say,   as  has been found numerically in general
and must be  at least approximately the case for a strict Laplace  resonance 
for which $W_{2,3}=2W_{2,1}/(1-W_{2,1})$  (see equation~[\ref{LAPR}]). In addition, we take
the planet masses to be comparable and $\sim m.$
Then the characteristic evolution expected from equations~(\ref{ELcon1})~and~(\ref{ELWcon}) 
can  be obtained from an equation of the form
$dW_0/dt \sim (m/M)^2/(t_cW_0^2),$ 
where we take $t_c$ to be the shortest circularization time $t_{e,i}.$
Thus:
\begin{equation}  W_0 \sim \left( \frac{m}{M} \right)^{2/3} 
\left( \frac{t-t_0}{3t_c} \right)^{1/3}, \label{Hscal}\end{equation}
where $t_0$ specifies an arbitrary origin of time.
We recall  that (\ref{Hscal})  applies to a two planet system by 
appropriately  setting $m_3=0$ as well as a three planet system.  
The time required to attain a departure $W_0$  is $t-t_0\sim 3 W_0^3 (M/m)^2 t_c$.
In general,
this is very
much longer than the  characteristic circularization time.

\end{appendix}


\begin{thebibliography}{}


\bibitem[2009]{Barnes}
Barnes R., Jackson B., Raymond S. N., West A. A., Greenberg R., 2009, ApJ, 695, 1006

\bibitem[2005]{Bonfils}
Bonfils X., Forveille T., Delfosse X., et al. 2005, A\&A, 443, L15

\bibitem[1993]{Bouvier}
Bouvier J.,  Cabrit S., Fernandez M., Martin E. L., Matthews J. M., 1993, A\&A, 272, 176

\bibitem[1961]{Brouwer}
Brouwer D., Clemence G. M., 1961, Methods of celestial mechanics, New York: Academic Press

\bibitem[2005]{Brunini}
Brunini A., Cionco R. G., 2005, Icarus, 177, 264

\bibitem[1966]{Goldreich}
Goldreich P., Soter S., 1966, Icarus, 5, 375

\bibitem[2007]{Ivanov}
Ivanov P. B., Papaloizou J. C. B., 2007, MNRAS, 376, 682

\bibitem[2008]{Kennedy}
Kennedy G. M., Kenyon S. J., 2008, ApJ, 682, 1264

\bibitem[2008]{Leger}
L\'eger A., Rouan D., Schneider J., et al., 2008,  astro-ph/0908.0241

\bibitem[1974]{Lynden}
Lynden--Bell D., Pringle J. E., 1974,  MNRAS, 168, 603

\bibitem[2009]{Mayor1}
Mayor M., Bonfils X., Forveille T., et al. 2009a,  astro-ph/0906.2780

\bibitem[2009]{Mayor2}
Mayor M., Udry S., Lovis C., et al. 2009b, A\&A, 493, 639

\bibitem[1999]{Murray}
Murray C. D., Dermott S. F., 1999, Solar System Dynamics (CUP), p.~254--255

\bibitem[2003]{Papaloizou1}
Papaloizou J.C.B., 2003, Cel. Mech. and Dynam. Astron., 87, 53

\bibitem[2000]{Papaloizou4}
Papaloizou J. C. B.,  Larwood J. D., 2000,  MNRAS, 315, 823

\bibitem[2005]{Papaloizou3}
Papaloizou  J.C.B., Szuszkiewicz E., 2005, MNRAS,  363, 153

\bibitem[2009a]{Papszusz}
 Papaloizou J.C.B., Szuszkiewicz  E., 2009, astro-ph:0911.1554
 
\bibitem[2001]{Papaloizou2}
Papaloizou J.C.B., Terquem  C., 2001, MNRAS, 325, 221

\bibitem[2006]{Paardekooper1}
Paardekooper S.--J, Mellema G., 2006, A\&A, 459, L17

\bibitem[2006]{Paardekooper2}
Paardekooper S.--J, Papaloizou J. C. B., 2008, A\&A, 485, 877

\bibitem[2006]{Paardekooper3}
Paardekooper S.--J, Papaloizou J. C. B., 2009, MNRAS, 394, 2283

\bibitem[1993]{Press}
Press W.~H., Teukolsky S.~A., Vetterling W.~T., Flannery B.~P., 1993,
Numerical Recipes in FORTRAN
(CUP)

\bibitem[2008]{Raymond}
Raymond S. N.,  Barnes R.,  Mandell A. M., 2008, MNRAS, 384, 663

\bibitem[2009]{LInIda}	
Schlaufman, K. C.,  Lin, D. N. C.,  Ida S., 2009, Apj, .691, 1322

\bibitem[1975]{Sinclair}
Sinclair  A. T., 1975, MNRAS, 171, 59


\bibitem[2002]{Tanaka}
Tanaka H., Takeuchi T., Ward W. R., 2002, ApJ, 565, 1257

\bibitem[2007]{Terquem}
Terquem C., Papaloizou J. C. B., 2007, ApJ, 654, 1110

\bibitem[2007]{Udry}
Udry S., Bonfils X., Delfosse X., et al. 2007, A\&A, 469, L43

\bibitem[1997]{Ward}
Ward W. R., 1997, Icarus, 126, 261

\bibitem[2009]{Zhou}
Zhou J.--L., 2009, astro-ph/0902.4086

\end{thebibliography}
\end{document}